\documentclass[fleqn,usenatbib]{mnras}

\usepackage{mathptmx}

\usepackage[T1]{fontenc}

\DeclareRobustCommand{\VAN}[3]{#2}
\let\VANthebibliography\thebibliography
\def\thebibliography{\DeclareRobustCommand{\VAN}[3]{##3}\VANthebibliography}

\usepackage{graphicx}	
\usepackage{amsmath}	
\usepackage{amssymb}	
\usepackage{hyperref}
\usepackage{xcolor}

\def\msunh{\,h^{-1}\rm{\,M_{\odot}}}
\def\msun{\rm{\,M_{\odot}}}

\def\kms{\rm{\,km\,s^{-1}}}

\definecolor{orange}{rgb}{1,0.5,0}

\definecolor{amethyst}{rgb}{0.6, 0.4, 0.8}
\definecolor{ao(english)}{rgb}{0.0, 0.5, 0.0}

\title[Halo cluster shapes]{Halo cluster shapes: Insights from simulated galaxies and ICL with prospects for weak lensing applications}
\author[Gonzalez et al.]{Elizabeth J. Gonzalez$^{1,2}$\thanks{E-mail: ejgonzalez@unc.edu.ar}, Cinthia Ragone-Figueroa$^{1,2,3}$, Carlos J. Donzelli$^{1,2}$, 
\newauthor
Mart\'in Makler$^{4,5}$, Diego Garc\'ia Lambas$^{1,2}$, Gian Luigi Granato$^{1,3,6}$   
 \\
$^{1}$ Instituto de Astronom\'{\i}a Te\'orica y Experimental (IATE-CONICET),
 Laprida 854, X5000BGR, C\'ordoba, Argentina.\\
$^{2}$ Observatorio Astron\'omico de C\'ordoba, Universidad Nacional de C\'ordoba, Laprida 854, X5000BGR, C\'ordoba, Argentina.\\
$^{3}$ INAF, Osservatorio Astronomico di Trieste, via Tiepolo 11, I-34131 Trieste, Italia\\
$^{4}$ International Center for Advanced Studies \& Instituto de Ciencias F\'isicas,  ECyT-UNSAM \& CONICET, 1650, Buenos Aires, Argentina\\
$^{5}$ Centro Brasileiro de Pesquisas F\'isicas, Rua Dr. Xavier Sigaud 150, CEP 22290-180, Rio de Janeiro, RJ, Brazil\\
$^{6}$ IFPU - Institute for Fundamental Physics of the Universe, Via Beirut 2, I-34014 Trieste, Italia 
}

\date{Accepted XXX. Received YYY; in original form ZZZ}

\pubyear{2020}

\begin{document}
\label{firstpage}
\pagerange{\pageref{firstpage}--\pageref{lastpage}}
\maketitle
\begin{abstract}
We present a detailed study of the shapes and alignments of different galaxy cluster components using hydrodynamical simulations. We compute shape parameters from the Dark Matter (DM) distribution, the galaxy members and the intra-cluster light (ICL). We assess how well the DM cluster shape can be constrained by means of the identified galaxy member positions and the ICL.
Further, we address the dilution factor introduced when estimating the cluster elongation using weak-lensing stacking techniques, which arises due to the misalignment between the total surface mass distribution and the distribution of luminous tracers. The dilution is computed considering the alignment between the DM and the Brightest Cluster Galaxy, the galaxy members and the ICL. Our study shows that distributions of galaxy members and ICL are less spherical than the DM component, although both are well aligned with the semi-major axis of the later. We find that the distribution of galaxy members hosted in more \textit{concentrated} subhalos is more elongated than the distribution of the DM. Moreover, these galaxies are better aligned with the dark matter component compared to the distribution of galaxies hosted in less \textit{concentrated} subhalos. We conclude that the positions of galaxy members can be used as suitable tracers to estimate the cluster surface density orientation, even when a low number of members is considered. Our results provide useful information for interpreting the constraints on the shapes of galaxy clusters in observational studies. 
\end{abstract}

\begin{keywords}
galaxies: clusters: general galaxies: clusters: intracluster medium (cosmology:) dark matter gravitational lensing: weak methods: numerical
\end{keywords}


\section{Introduction}

According to the hierarchical formation scenario, dark matter halos in which galaxies and galaxy systems reside,
evolve under the accretion and merger of smaller halos. Observational evidence as well as numerical simulations have shown that this accretion mainly occurs along preferential directions, traced by the filamentary structure of the cosmic web \citep[e.g.][]{Guo2015,Tempel2015,Gonzalez2016}. Therefore, halos are not expected to be spherical but to have triaxial shapes and appear to be elliptical in projection, which is confirmed by numerical simulations
\citep[e.g.][]{Dubinski1991,Warren1992,Cole1996,Jing2002,Bailin2005,Hopkins2005,Kasun2005,Allgood2006,Paz2006,Bett2007,Munoz2011,Schneider2012,Despali2013,Velliscig2015,Vega-Ferrero2017}. 

Shape measurements of dark matter halos constitute a test of the current cosmological $\Lambda$CDM
scenario, since they are sensitive to the initial density field and halo assembly
\citep{Kawahara2010,Sereno2018}. Determining halo shapes
can also serve as a probe of the fundamental particle nature of dark matter. 
For instance, in simulations that include self-interacting dark matter, halos become rounder towards the centre as the 
cross-section of the dark matter particle increases \citep{Robertson2019}. Moreover, constraining the shape of the halos provides information regarding their formation history \citep{ragone2010, Lau2020}. Simulations predict that halos are less spherical and
more prolate with increasing mass and redshift due to the hierarchical formation process. More massive halos form later and their shapes are more affected by the last major merger event
\citep{vanHaarlem1997,Colberg2000,Vitvitska2002,Porciani2002}. Despite the correlation between the halo mass and shape \citep[e.g.][]{Shaw2006,Allgood2006,Velliscig2015,Vega-Ferrero2017}, a large scatter of the shape parameters at a fixed halo mass is present. The scatter likely originates from the different formation histories. Indeed, the time elapsed since the last major merger event, or the dynamical age, plays a significant role in shaping the halo \citep[e.g.][]{ragone2010, Chen2020}.

Deriving observational estimates of cluster dark matter halos shapes is challenging. One of the most promising observational techniques is gravitational lensing. Weak and strong lensing analyses
of individual galaxy clusters provide direct shape measurements 
of the projected mass distribution \citep[e.g.][]{Richard2010,Oguri2010,Umetsu2018,Okabe2020}.
However, these studies are mainly restricted
to massive clusters ($ \gtrsim 10^{15} M_\odot$) since the shape
measurements strongly depend on the lensing signal. A possibility to overcome this limitation is to use weak lensing stacking techniques, which combine a sample of clusters with a similar property, such as the galaxy member richness, to increase the lensing signal-to-noise ratio. These techniques are usually applied to calibrate the relation between halo masses and cluster observables \citep[e.g.][]{vonderLinden2014,Applegate2014,Hoekstra2015,McClintock2019,Murata2019,Pereira2020}. If the lensing signal is combined considering the orientation of the cluster projected mass distribution, the  resulting stacked mass density can be modelled to derive the projected ellipticity of the total mass surface density \citep{Evans2009,Oguri2010,Clampitt2016,vanUitert2017,shin2018,Gonzalez2021}. However, since the true orientation of individual clusters is unknown, observable proxies have to be considered. The proxies most often used to trace the halo orientation are the major semi-axis of the brightest cluster galaxy (BCG) and the distribution of galaxy cluster members. 

One straightforward observational approach to estimate the shape of the dark matter cluster halos uses the galaxy members' position
to derive the projected semi-axis ratio. Projected shapes derived from this technique are biased to more elongated distributions due to the noise bias introduced by the low number of considered tracers. 
Additionally, shape computations can also be affected by interloper galaxies and the specifics 
of the cluster membership assignment. Despite the aforementioned caveats, cluster shapes estimated using the galaxy member positions have yield results in agreement with simulations \citep[e.g.][]{huang2016,shin2018}. 

Another potential luminous tracer of the total dark matter distribution in galaxy clusters is the intracluster light (hereafter ICL). This light arises from stars that are not gravitationally bound to any galaxy, a population probably originated from tidal stripping effects and mergers of the cluster members \citep[e.g.,][]{Gonzalez2005,Zibetti2005,Murante2007,Mihos2017,Zhang2019b,Contini2019,Montes2019}. From the observational side there are some indications that the ICL could be aligned with the host cluster, as traced by the galaxy distribution, and that this alignment would be tighter than the already known BCG-Cluster one \citep{Kluge2021}. Moreover, recent studies based both on observations and simulations have explored the relation between the ICL and the global dark matter distribution, hinting at a strong link between these two cluster components, although with observed discrepancies in the radial distribution \citep[e.g.,][]{Pillepich2018,montestrujillo2019,Alonso-Asensio2020,Sampaio-Santos2021}. These results encourage us to asses the use of this luminous tracer to further constrain cluster halo shapes. 

In this work we analyse the shape distributions of the different cluster components using a set of simulated clusters that include baryonic physics. 
Given that clusters are mainly dominated by the dark matter component, the general trends between the cluster mass, redshift and shapes can be obtained from gravity only simulations without the need of baryon physics \citep[e.g.][]{ragone2007,ragone2010,Despali2017}. Nevertheless, baryon effects have a significant impact on modelling the cluster shapes, mainly at the inner regions \citep[roughly at $20\%$ of the virial radius,][]{Cataldi2020}. In general, when baryon physics are incorporated, halos tend to be rounder when approaching to the central region \citep{Velliscig2015,Suto2017,Chua2019}. 

The present analysis includes
the study of the dark matter mass distribution of simulated galaxy clusters and its relation to the stellar distribution, particularly the ICL and the galaxy members. We also investigate the bias introduced in the elongation estimator from gravitational lensing when considering these tracers to align the clusters to perform stacking techniques. 
The paper is organised as follows. In Sec. \ref{sec:numsim} we describe the main characteristics of the simulated clusters. In Sec. \ref{sec:params} we define the shape parameters and detail how we compute these parameters for each tracer. We also summarise the general trends observed between the dark matter and stellar shape parameters as a function of 
the distance to the cluster centre. In Sec. \ref{sec:results} we present an analysis of how luminous tracers can be used to assess the cluster shape in observations and discuss possible biases introduced  from using these tracers to determine the orientation of the projected mass distribution of the cluster. To this end, we consider the shape parameters derived according to the galaxy members and the stellar distributions and the fitted ICL. Considering the estimated position angles according to these tracers, we also present
the estimated biases introduced in the weak lensing stacking studies encoded in the so-called dilution factor. Finally, we summarise our results and conclude in Sec. \ref{sec:conclusion}.

\section{Simulated clusters}
\label{sec:numsim}
\subsection{Numerical Simulations}
\label{simprop}
The present work is based on the set of hydrodynamical simulations
described in \cite{ragone2018}, which introduce an improved AGN feedback with respect to those presented in \cite{ragone2013}. We refer the reader to the above two papers for the numerical or technical specifications. The first paper shows that these simulations reproduce quite well the BCG mass evolution derived from observations. Additionally, the same set of simulations has recently been used to study the BCG-Cluster alignment evolution during the last 10 Gyr \citep{Ragone2020} and the persistence of this alignment for off-centre BCGs \citep{depropris2021}. In the following, we summarise the features that are more relevant to the present study.

The set of  initial conditions comprises 29 zoomed-in Lagrangian regions which we evolve with a custom version of the {\footnotesize GADGET-3} code~\cite[][]{springel2005}. These regions have been selected from a parent gravity-only simulation of a 1 $h^{-1}$Gpc box surrounding the 24 most massive dark matter (DM) halos. They all have masses\footnote{$M_\Delta$ is the mass enclosed by a sphere whose mean density is $\Delta$ times the critical density of the Universe at the considered redshift. The radius of this sphere is dubbed $R_\Delta$} $M_{200} \gtrsim 1.1 \times 10^{15}\, \msun$. In addition 5 less massive haloes with masses $1.4 \times 10^{14} \lesssim M_{200} \lesssim 6.8 \times 10^{14}\, \msun$ are selected randomly. Each region was re-simulated at higher resolution and taking into account hydrodynamics and sub-resolution baryonic processes. 
The adopted cosmology is defined by the following parameter values: $\Omega_{\rm{m}} = 0.24$, $\Omega_{\rm{b}} = 0.04$, $n_{\rm{s}}=0.96$, $\sigma_8 =0.8$ and $H_0=72\,\kms$\,Mpc$^{-1}$. The mass resolution is  $m_{\rm{DM}} = 8.47\times10^8 \, \msunh$ for the DM and $m_{\rm{gas}} =1.53\times10^8\, \msunh$ for the gas (initial). When computing gravitational interactions, a Plummer-equivalent softening length of $\epsilon = 5.6\, h^{-1}$\,kpc is adopted for gas  particles, while $\epsilon = 3\, h^{-1}$\,kpc for black hole and star particles. As for DM particles, the softening length is set to $13.8\, h^{-1}$\,kpc at $z>2$ and later on to $5.6\, h^{-1}$\,kpc.  

The simulations include sub-resolution prescriptions for several baryonic processes.  Details on the adopted implementation of cooling, star formation, and associated feedback, are given in \cite{ragone2013}.
Metallicity dependent cooling is implemented following \cite{wiersma2009}. Metal enrichment is treated as in \cite{tornatore2007}. 
Each spawned stellar particle in the simulation represents a Single Stellar Population (SSP) with a Initial Mass Function (IMF) of \cite{chabrier2003}. 
The AGN feedback model is described in Appendix A of \cite{ragone2013}, with a few modifications discussed in Section 2 of \cite{ragone2018}, meant to improve the spatial association of the SMBH particles with the stellar system in which they are seeded. A stable association is demanded to maximise the effect of AGN feedback in limiting the growth of stellar mass. 

The re-simulated volumes are chosen to ensure that by $z=0$ no dark-matter particles coming from the low-resolution region fall within 5 virial radii from the target cluster centre.  For this reason, besides the central cluster,  more clusters are present in the same Lagrangian region. In this particular work, we selected in each region those clusters that at $z=0$ have at least 10 galaxies within $R_{200}$ with stellar masses higher than $10^{10}\msun$ (see \ref{subsec:members} for the galaxy identification). This selection criterion leads us to a sample of 72 clusters at $z=0$ whose $M_{200}$ distribution ranges from $10^{14}\msun$ to $4 \times 10^{15}\msun$ with a median value of $\sim 4\times 10^{14}\msun$. The main clusters (MCs) of each Lagrangian region have instead a mass distribution with a median value of $\sim 1.6 \times 10^{15}\msun$. For these particular 29 MCs we  have determinations of the cluster formation time, defined as the time at which the cluster assembled half of its final $M_{200}$ mass.

Besides the 3D properties, each cluster has also determinations of the projected 2D properties computed along the three Cartesian axes. The later takes into account all the matter (dark-matter, stellar particles, galaxies) within the re-simulated region ($\sim 40\, \text{Mpc}$) that 
falls inside the projected radius within which the determination is done. 
In order to provide observational constrains, we will focus the main body of the paper to study the projected sample, with a total of 3x72=216 clusters, out of which 3x29=87 are MCs. 


\subsection{Galaxy members and centre definition} 
\label{subsec:members}

 We use  the {\footnotesize SUBFIND} subhalo finder algorithm \citep{springel2001, dolag2009} to identify the subhaloes (or galaxies) within each cluster. {\footnotesize SUBFIND} uses the particles of the main FOF halo to determine saddle points in the density field. Particles lying inside borders defined by the spatial position of saddle points are grouped together. Finally, an unbinding procedure discards high speed particles, not gravitationally bound to the structure. 
The main subhalo in each cluster includes the BCG and all the particles that are not bounded to any other subhalo, i.e. the intra-cluster stars. The BCG centre coincides with the cluster halo centre and is given by the position of the particle in the main {\footnotesize SUBFIND} subhalo that has the minimum gravitational potential. 
 
In the present analysis, we only consider in 3D and in projection as galaxy members those subhalos with a stellar mass higher than $10^{10}M_\odot$. In the 2D analysis, we include as galaxy members all the subhalos identified within the cylinder along the re-simulated region. This procedure partially mimics the presence of interlopers (foreground and background galaxies) in observations. 

To explore the shape distribution of subsamples of galaxies with different characteristics, we classify all the subhalos within the main halo $R_{200}$ as \textit{concentrated} or \textit{extended} depending on the radius that encloses half of the subhalo total mass, $R_{1/2}$. We define \textit{concentrated} (\textit{extended}) galaxies as those subhaloes whose $R_{1/2}$ are lower (higher) than the median  $R_{1/2}$ in the same cluster mass bin. The resolution of the present simulations is not sufficient to safely capture the galaxy morphological type. However, it is conceivable that more \textit{concentrated} halos are those that would tend to host early-type galaxies in more realistic simulations. We inspect the radial distribution of both galaxy subsamples obtaining that more \textit{concentrated} galaxies populate preferentially the central region of the clusters as opposed to the more \textit{extended} galaxy subsample. This result is in agreement with previous studies showing that at fixed masses, the subhalo concentration significantly increases towards the host halo centre
\citep{Diemand2008,Pieri2011,Moline2017,wang2018}.

\subsection{Relaxation classification}
\label{subsec:relax}
\begin{figure}
\centering
    \includegraphics[scale=0.6]{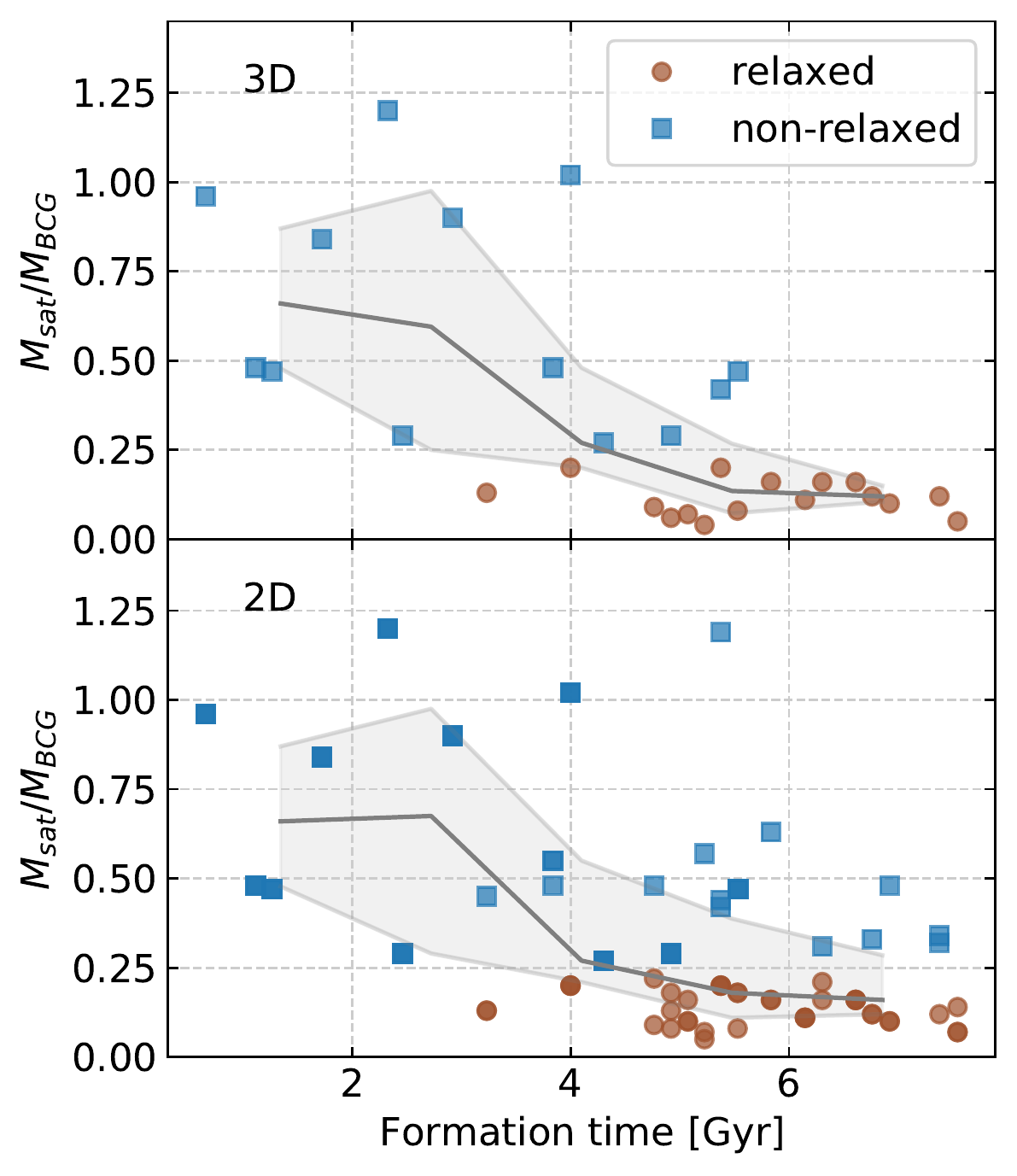}
    \caption{Relaxation halo proxy, given by the mass ratio ($M_\text{sat}/M_\text{BCG}$),  plotted against the cluster formation time for the 29 MCs in 3D (upper panel) and the 87 in 2D projections (lower panel).  
    Blue squares (red dots) represent the clusters classified as non-relaxed (relaxed). Some overlap between the two groups occurs in the 2D panel.
    Solid grey line corresponds to the median values computed in formation time bins and the shadow region encloses the 25th-75th percentiles. The clusters relaxation proxy $M_\text{sat}/M_\text{BCG}$ correlates well with the formation time of clusters  }
    \label{fig:proxies}
\end{figure}

To explore the relation between the shape parameters and the relaxation and dynamical age of the halos, we classify the clusters as relaxed and non-relaxed according to an observationally related cluster property: the mass ratio between the most massive satellite galaxy and the BCG, $M_\text{sat}/M_\text{BCG}$. Here $M_\text{sat}$ is the mass of the highest mass subhalo located within a radius of $0.5 R_{200}$ while $M_\text{BCG}$ is the BCG mass computed as the sum of the stellar particles within $0.1R_{500}$. This mass ratio is expected to be related to the relaxation condition of the galaxy clusters, with lower values indicating a more relaxed status \citep[e.g.][]{Zhoolideh2020}. We compute this parameter for each cluster in 3D and in the three corresponding 2D projections. We classify the clusters according to the median value of the mass ratio distribution in 2D, considering as relaxed (non-relaxed) those clusters with $M_\text{sat}/M_\text{BCG} \leq 0.22$  ($M_\text{sat}/M_\text{BCG} > 0.22$). Assuming a constant mass to light ratio, this value corresponds to a magnitude gap of $\sim 1.6$, not far from the conventional value of $2$ adopted to define fossil groups \citep{Ponman1994}.
Consequently, $61\%$ of the clusters are classified as relaxed in 3D, and  $52\%$ in 2D. 

 To assess the ability of our proxy to estimate the relaxation state of clusters, we compare it with the cluster formation time for the MCs. Fig. \ref{fig:proxies} shows that there is a good correlation between the formation time and the $M_\text{sat}/M_\text{BCG}$ mass ratio, indicating that clusters classified as relaxed by our observational proxy criterion tend to have higher formation times. However, in projection some early-formed clusters are classified as non-relaxed.
 
 We also adopt two other usual observable relaxation proxies to classify the clusters: the distance offset between the BCG location and the cluster centre computed using satellite positions, and the median satellite bulk velocity with respect to the BCG. The observed differences in the shape distributions between relaxed and non-relaxed clusters in the next sections are in agreement regardless the proxy considered. 
 We finally adopt the mass ratio for this classification since this parameter shows a tight relation with the cluster formation times and it does not depend on spectroscopic information as required by the velocity difference proxy. Moreover, we find that the mass ratio shows a better agreement between the 3D and the 2D classification than those considering distance offset and the velocity difference. 

\section{Shape parameters of the cluster components}
\label{sec:params}

\begin{table*}
 \caption{Summary of the methodology applied to derive the shape parameters for the different cluster components within the different radii.}
    \centering
    \begin{tabular}{c c c c c c c c c}
    \hline
    \hline
Cluster sample & Shape parameters & Tracer &   \multicolumn{3}{c}{BCG region} & \multicolumn{3}{c}{ICM}  \\
 \hline
Total sample &Inertial  & DM particles & $R <  30$kpc & $R < 50$kpc & $R < 0.1R_{500}$ & $R < R_{1000}$ & $R < R_{500}$ & $R < R_{200}$ \\
(72 clusters) & & Star particles &  $R < 30$kpc & $R < 50$kpc & $R < 0.1R_{500}$ & $R < R_{1000}$ & $R < R_{500}$ & $R < R_{200}$ \\
& & All galaxies$^{(*)}$  &  & - &   & $R < R_{1000}$ & $R < R_{500}$ & $R < R_{200}$ \\
& & \textit{Extended} galaxies$^{(**)}$  &  &  & -  &  &  & $R < R_{200}$ \\
& & \textit{Concentrated} galaxies$^{(**)}$  &  &  & -  &  &  & $R < R_{200}$ \\
\hline
MCs & Isocontours & Surface DM density  & \multicolumn{3}{c}{$R < 0.1R_{500}$} & \multicolumn{3}{c}{$0.1R_{500} < R \leq R_\text{LIM}$} \\
(29 clusters) & & Surface brightness density  & \multicolumn{3}{c}{$R < 0.1R_{500}$} & \multicolumn{3}{c}{$0.1R_{500} < R \leq R_\text{LIM}$} \\
\hline
\end{tabular}
\begin{flushleft}
$(*)$ Galaxy member shape parameters are obtained for clusters with more than 10 identified members in 3D, leading to a total number of 64 and 71 clusters with shape estimates within $R_{1000}$
and $R_{500}$, respectively. \\
$(**)$ \textit{Concentrated} and \textit{extended} galaxy distribution shapes are obtained for the 59 clusters with more than 20 identified members in 3D.
\end{flushleft}
  
    \label{tab:shapes}
\end{table*}

This section introduces the shape parameters for the different cluster components or, as dubbed in this work, tracers. The tracers we analyse are star particles, galaxy members and dark-matter (DM) particles. We study the shape parameters at different cluster-centric regions. The BCG region is defined by the DM and stars particles within $0.1R_{500}$, while beyond this radius begins what we consider the intracluster cluster medium (ICM). 
The choice of $0.1R_{500}$ as the BCG boundary, is motivated by the results presented by \cite{ragone2018}, who obtain that roughly at this radius the surface brightness drops to $\mu_V\sim24 {\rm~mag~arcsec^{-2}}$, which is a classical value adopted to define an observational galaxy limit \citep{devaucou1991}.

We derive shape parameters of the cluster components by using two different approaches: \begin{itemize}
\vspace{-\topsep}
    \item \textbf{Inertial shape parameters:} derived from the diagonalization of the shape tensor using the position of cluster galaxies, stars and DM particles with respect
    to the cluster centre within different radii. The computation of cluster shape parameters using galaxies is performed only if at least 10 galaxies are present within the implicated radius. These parameters are obtained for the total sample of clusters using the 3D tracer locations as well as for the three projected directions.
    \item \textbf{Isocontours shape parameters:} obtained from the best fitting ellipses of the surface brightness isophotes or the surface DM isodensity contours.  We compute them for the 29 MCs, projected perpendicularly to the three Cartesian axes. 
\end{itemize}
In Table \ref{tab:shapes} we detail the tracers considered and the cluster region at which the inertial and isocontour shape parameters are constrained. In the next subsections we describe how these parameters are derived. 

\subsection{Inertial Shape Parameters}
\label{subsec:inertial}

\subsubsection{Shape definition}

Inertial shape parameters for the different tracers of the galaxy clusters are computed using the position of the DM particles, stars and galaxy members as detailed in Appendix \ref{A:shape}. Projected shapes are characterised by the derived semi-axis ratio, $q$, and the position angle of the semi-major axis (SMA), used for computing the misalignment angle between the different tracers, $\theta$.
For all the mass tracers, inertial tensors are computed within three overdensity radii $R_\Delta$, with $\Delta = 1000, 500, 200$, related to the region that corresponds to the ICM. We also consider the stellar and the DM particle distributions within $30$kpc, $50$kpc and $0.1R_{500}$, which are related to the BCG region. Table \ref{tab:radius} reports the median values of the different radii for the total sample. 

In the case of the galaxy distribution, shape parameters are derived only if the number of galaxies within each considered radius is at least 10 in 3D, reducing the number of clusters with this information to 64 and 71 within $R_{1000}$ and $R_{500}$, respectively. Also, we derive cluster shapes considering \textit{extended}/\textit{concentrated} galaxies for the 59 clusters that have more than 20 total identified galaxies in 3D.  

\begin{table}
 \caption{Median values of the radius at which shape parameters are computed for the total sample of clusters. First three rows are related to the BCG region}
    \centering
    \begin{tabular}{c c c}
    \hline
    \hline
Definition &   $R$ & $R/R_{200}$  \\
 &    kpc & \\
    \hline
  -    &    30    &   0.02  \\
  -    &    50    &   0.03  \\
  $0.1R_{500}$    &    95    &   0.06  \\
  \hline
  $R_{1000}$   &    668   &   0.45  \\
  $R_{500}$    &    948   &   0.64  \\
  $R_{200}$    &    1452  &   1.00  \\
  \hline
  \end{tabular}
  \begin{flushleft}
\textbf{Columns:} (1) Radius definition according to the enclosed overdensity; (2) median values of the defined radius; (3) scaled median values according to $R_{200}$.
\end{flushleft}
    \label{tab:radius}
\end{table}

In the case of the DM distribution, we evaluated the impact of considering or not the particles associated with each subhalo in the shape determinations (see Appendix \ref{A:HvsDM}).  We found no significant differences besides a slightly more spherical shape when excluding particles associated with subhaloes. However, since the observational lensing shape parameters refer to the total matter distribution, we include sub-halo particles in the following analysis of our simulations. 

\subsubsection{General trends of dark matter and stellar distribution shape parameters}

Here we present the general trends obtained for the shapes of the DM and stellar distribution components. Overall, these results are in agreement with previous studies based on numerical simulations \citep[e.g.,][]{Velliscig2015,Vega-Ferrero2017,Despali2017,Chira2020}. We briefly comment our general results, and provide a more detailed analysis as complementary material in Appendices \ref{A:SP} and \ref{A:stars}.

According to our analysis, and in agreement with previous studies  including baryon physics, the DM distribution of galaxy clusters is rounder at the inner radii, both in 3D and in 2D projections. These trends are also present for the stellar distribution. The dark matter distribution is less spherical for higher mass clusters at all radii, especially when considering relaxed clusters. On the contrary, the stellar distribution of high and low mass relaxed clusters features similar median sphericity. As for the mass orientation, in relaxed clusters we find that the DM particles are well aligned at different cluster-centric regions.
%

Compared to the dark matter particle distribution, shapes derived from the stellar component are more prolate and less spherical at all radii. High-mass clusters show that both stellar and dark matter distributions are rounder within $30$kpc than in low-mass clusters. On the other hand, sphericity values of the stellar distribution within the same scaled radius $0.1R_{500}$ are similar for the high and the low-mass clusters. This result indicates that BCGs shapes are similar within the same scaled radius for the whole cluster mass range.

\begin{figure}
    \includegraphics[scale=0.6]{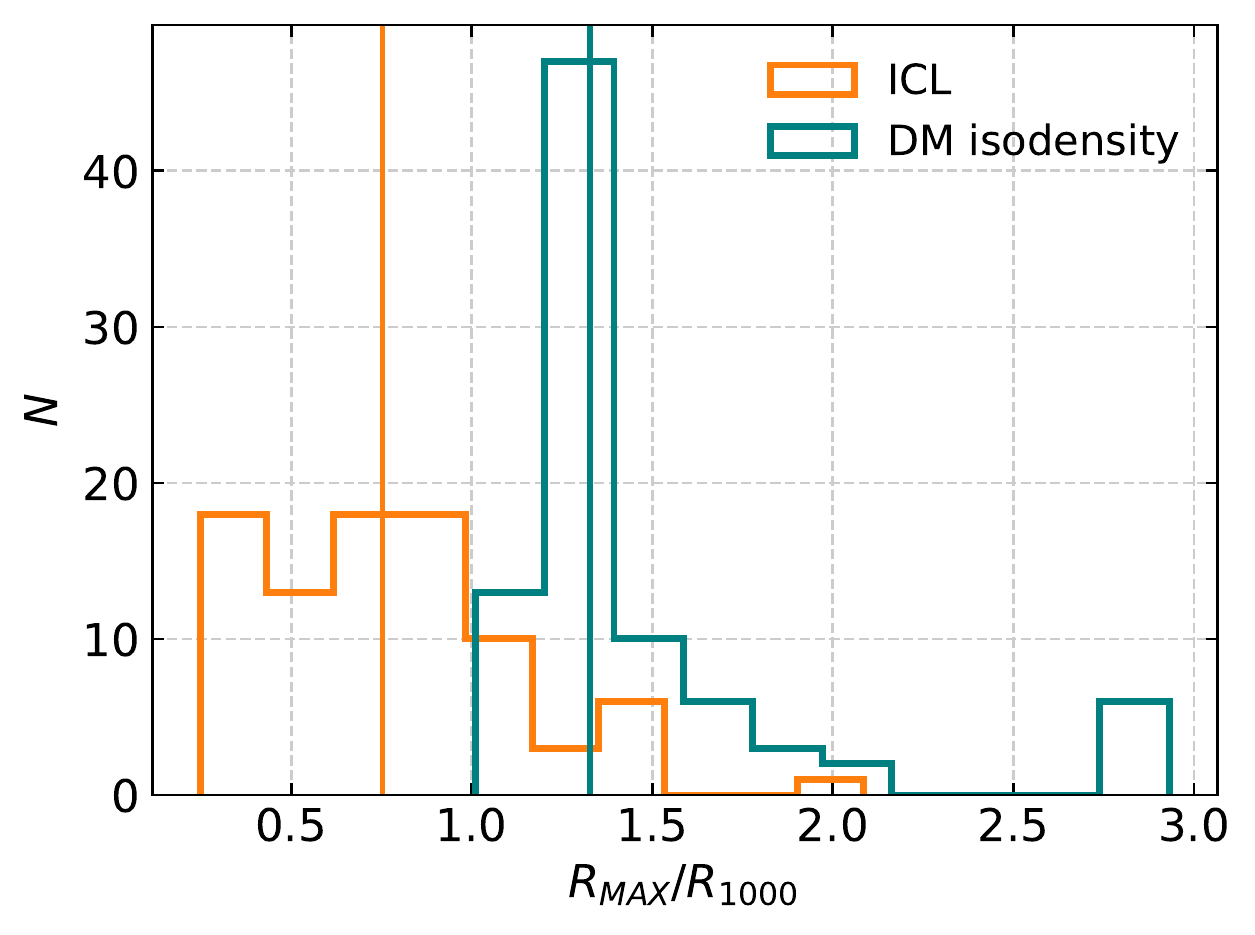}
    \caption{Distribution of the maximum SMA up to which the routine ELLIPSE is able to find a fit, $R_\text{MAX}$, scaled by $R_{1000}$. In orange we show the values for the ICL while in green for the DM. Vertical lines correspond to the median values.}
    \label{fig:histrmax}
\end{figure}

\begin{figure*}
    \includegraphics[width=5.8cm]{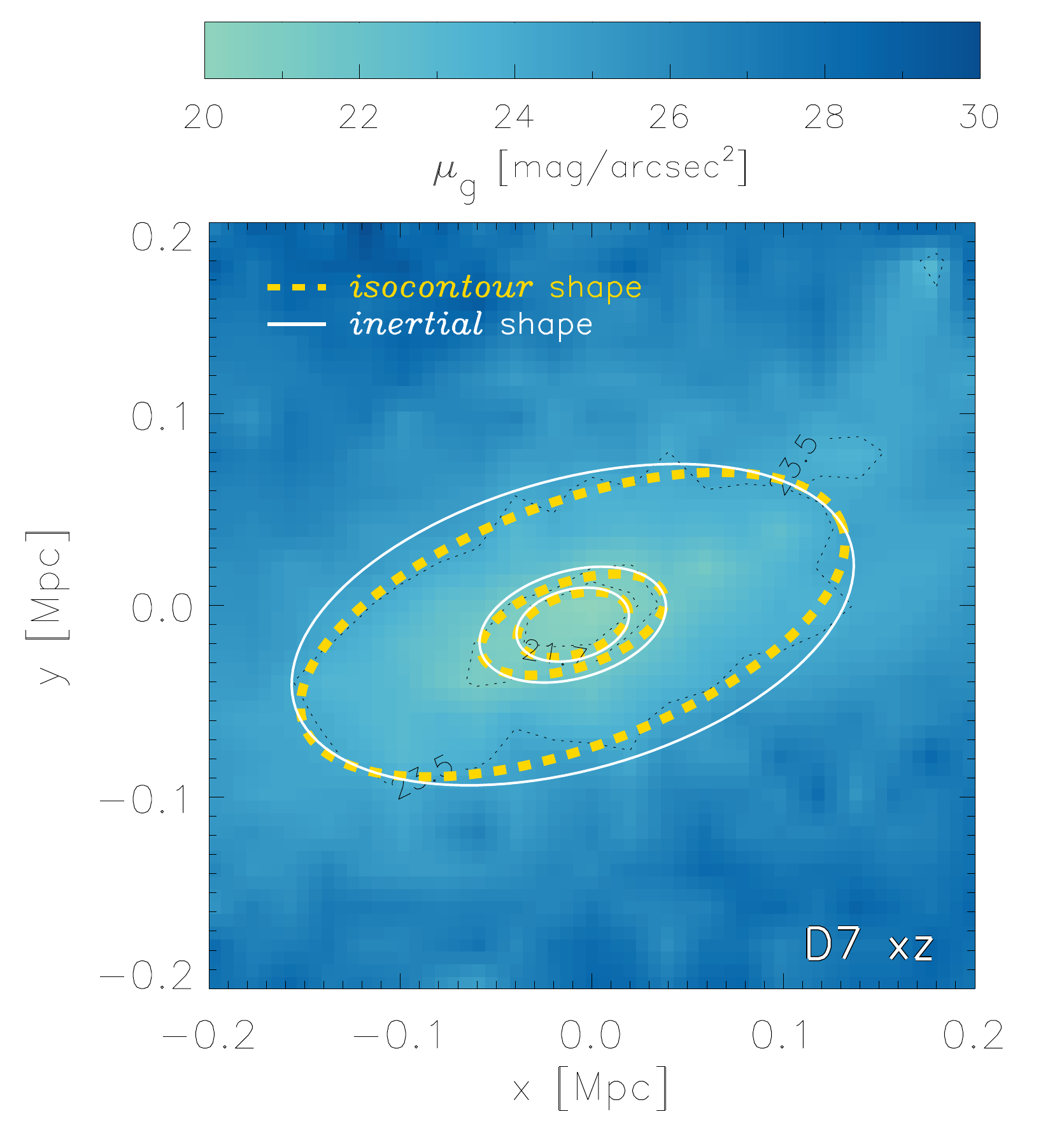}
    \includegraphics[width=5.8cm]{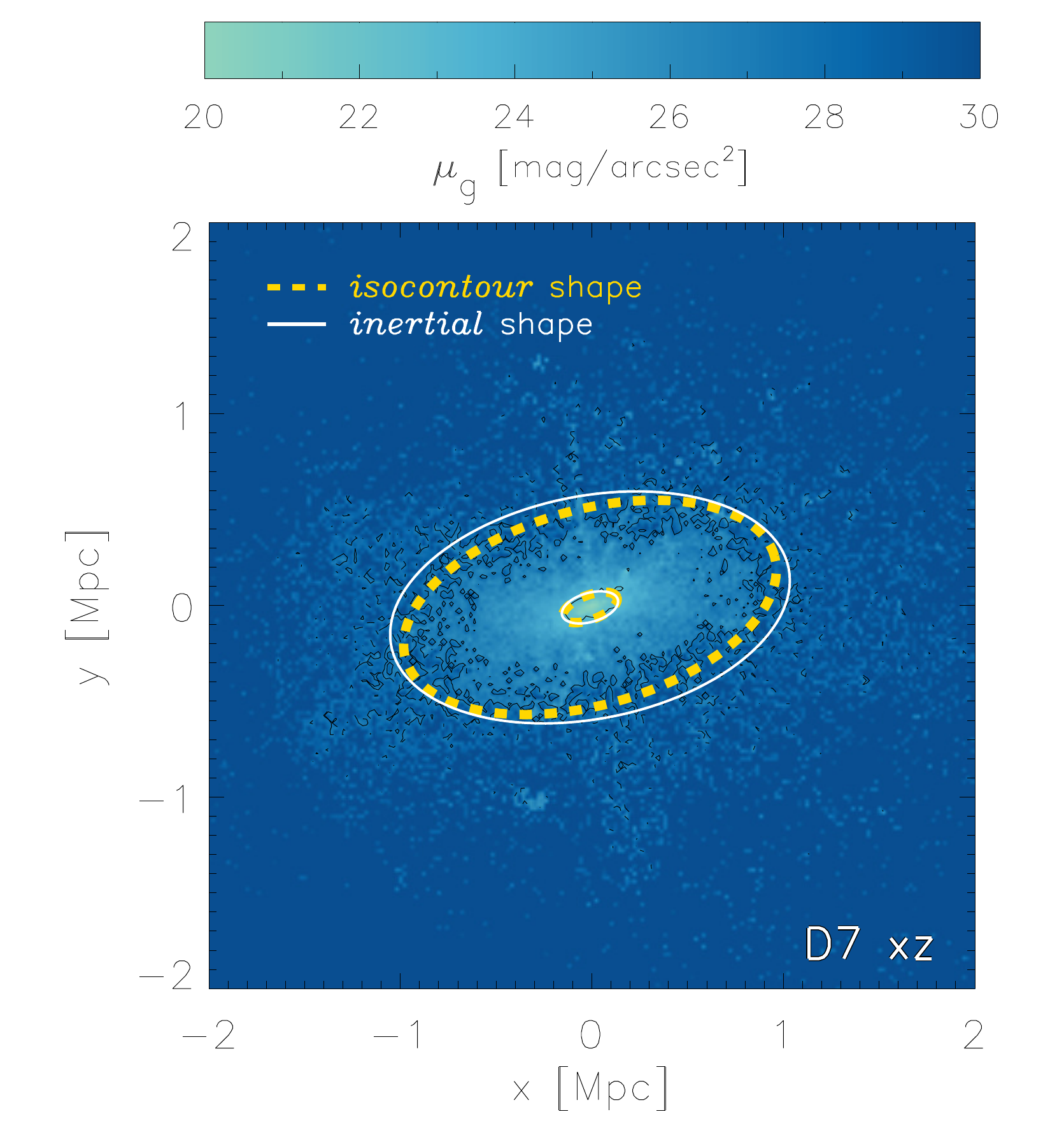}
    \includegraphics[width=5.8cm]{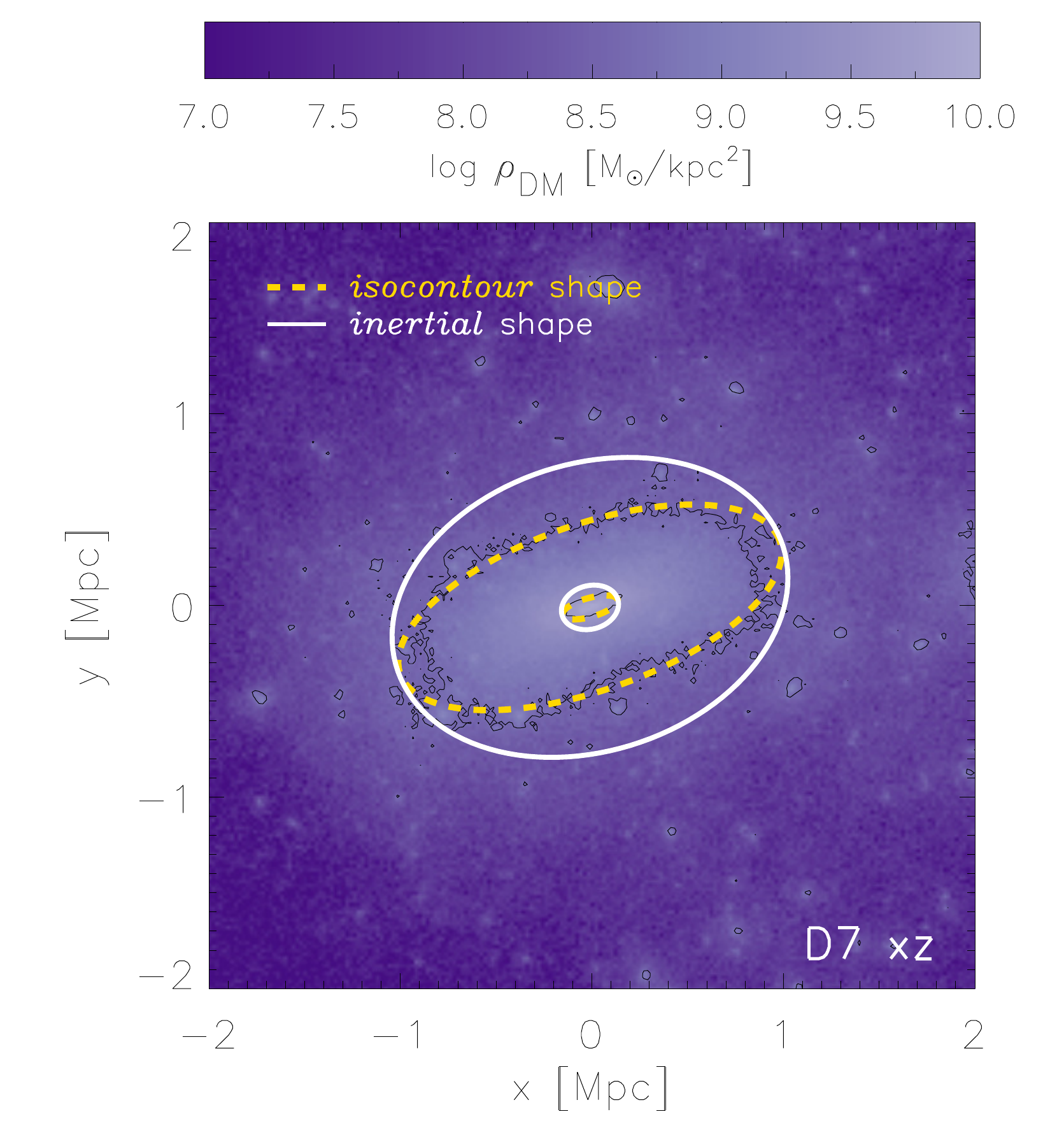}
    \includegraphics[width=5.8cm]{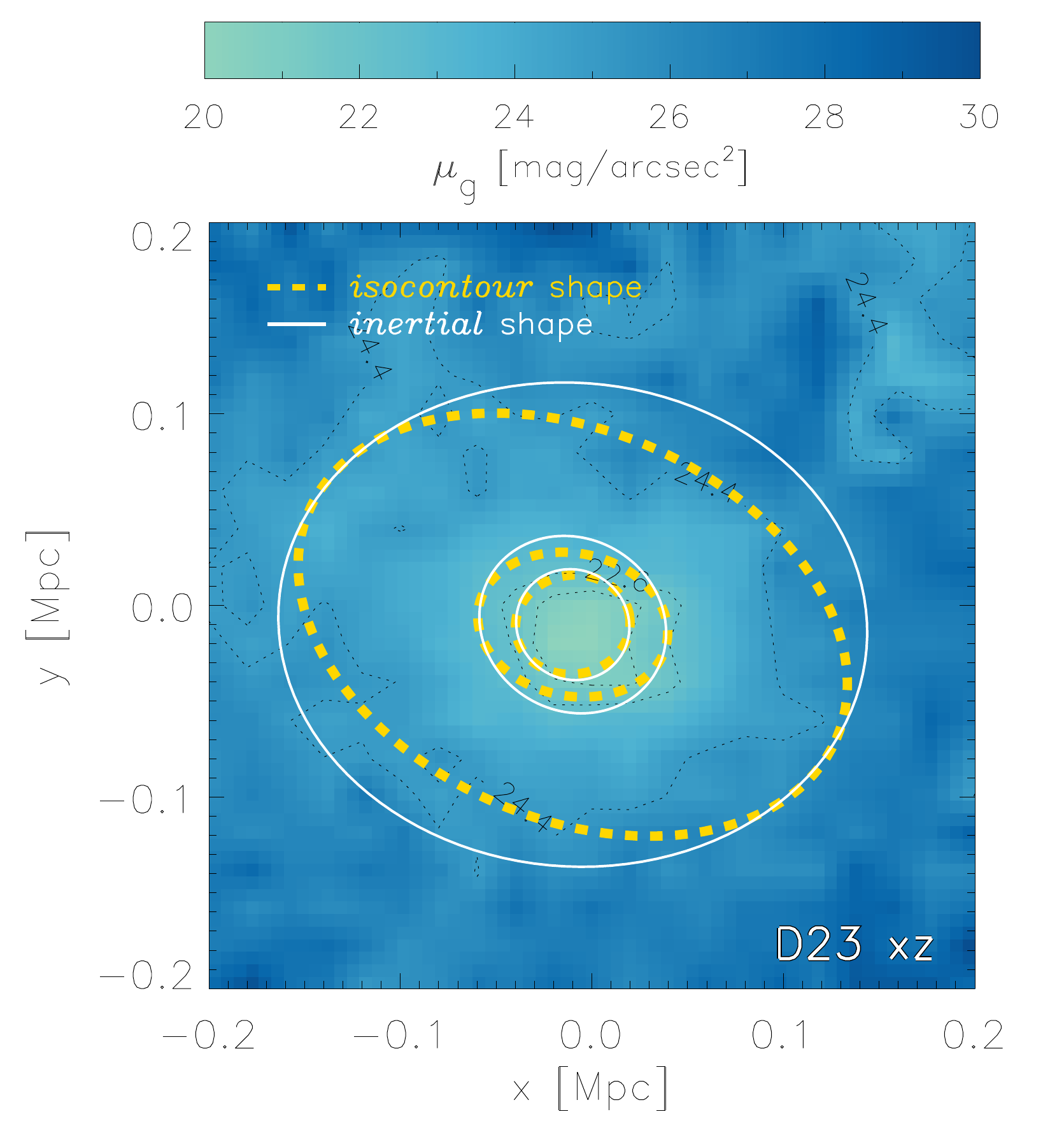}
    \includegraphics[width=5.8cm]{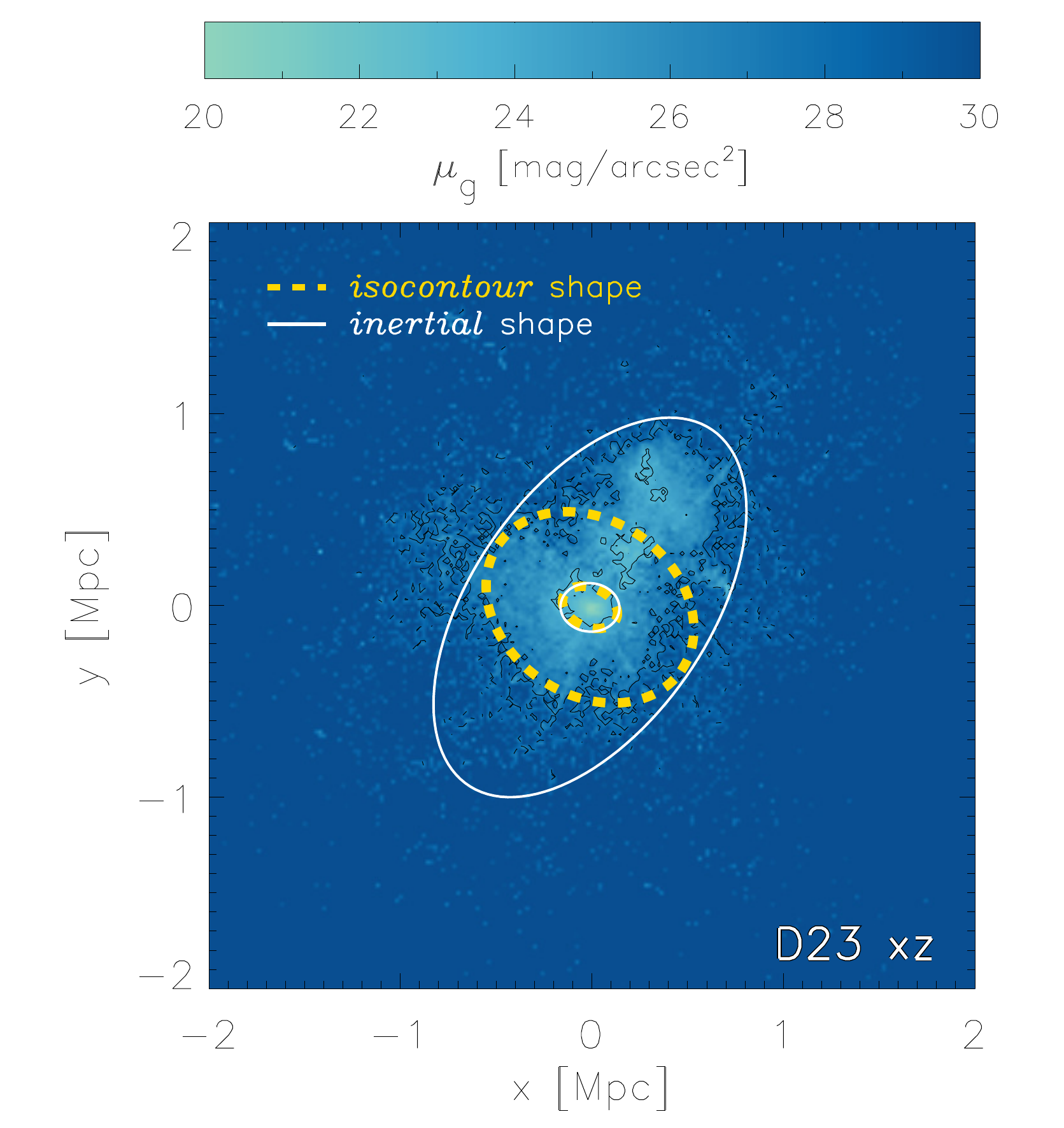}
    \includegraphics[width=5.8cm]{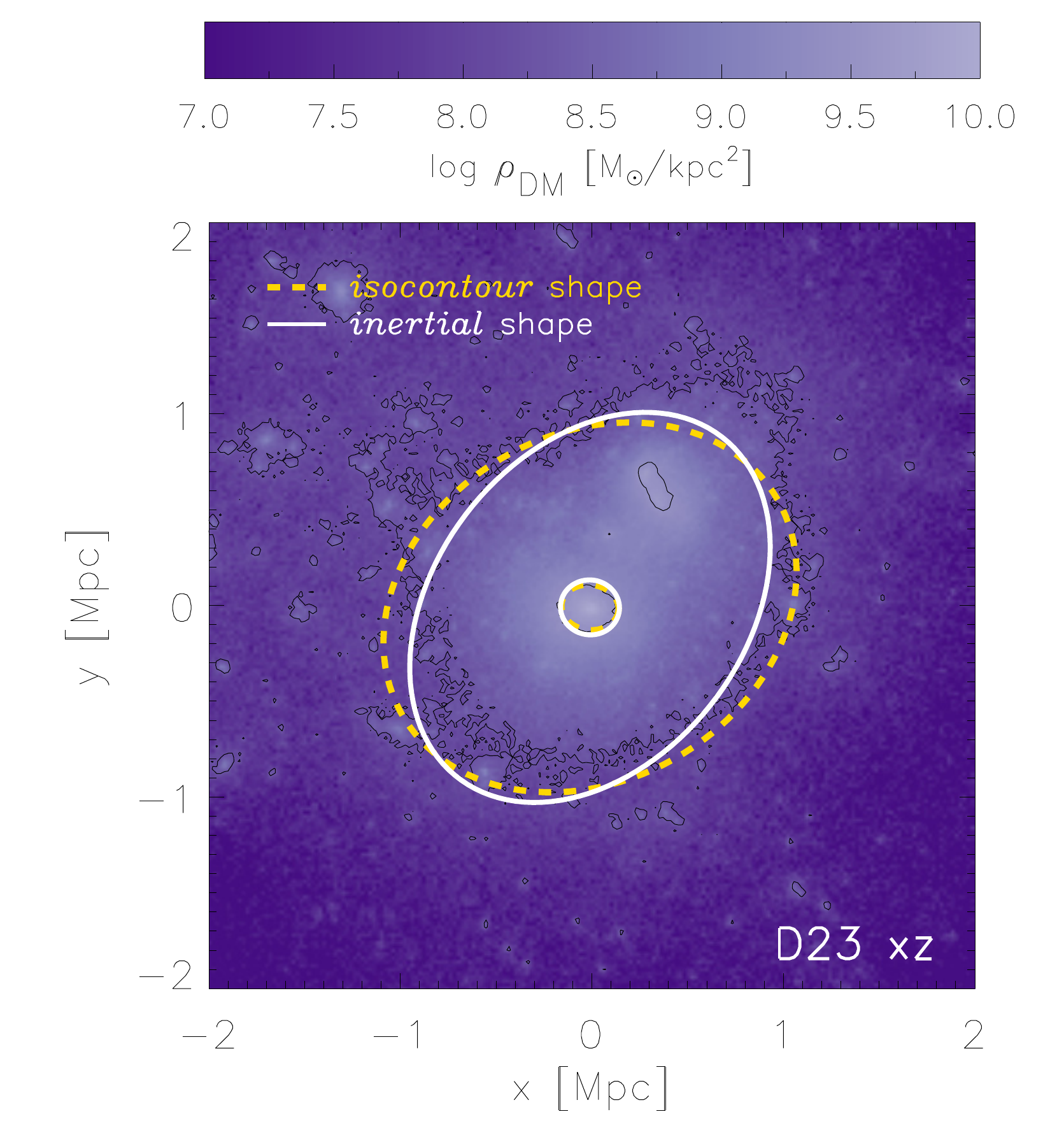}
    \caption{A schematic view of the shape and orientation angle determinations for an early and late formed cluster is presented in the top and bottom panels, respectively. Left and middle panels show  the surface brightness maps of the BCG and the BCG+ICL, respectively. 
    The right panels depicts instead the DM density distribution. In all the panels solid line ellipse parameters are obtained from the shape tensor selecting stars or DM particles, as the case may be, within different radii. Dashed line ellipses are instead the best fits returned by the rutine ELLIPSE for the same radii. The corresponding isophotes or isodensity contours are also shown in black. 
    In the case of the BCG we show elipses with SMAs of 30kpc (internal), 50kpc (middle) and 0.1$R_{500}$ (external). In the cluster scale we depict 0.1$R_{500}$ (internal) and $R_{1000}$ (external). Notice however that in the case of the ICL for the late-formed cluster (middle bottom) the {\it isocontour} shape determination does not reach up to $R_{1000}$, the proximity of a merging halo is apparent. 
    }
    \label{fig:bcgmaps}
\end{figure*}

\subsection{Isocontour Shape Parameters}

\subsubsection{Construction of Surface Brightness and DM density Maps}
\label{subsec:iclmod}
Mock surface brightness maps of the simulated clusters are generated only for the MCs.
As mentioned in Section \ref{simprop}, each stellar particle in the simulation represents a SSP. The assumed IMF, together with its age (time since spawning), metallicity and initial mass determine the particle spectral energy distribution (SED). We adopt the SSP templates by \cite{bruzual2003}. SDSS g-band filter is then applied to this SED to calculate the g-band luminosity which is smoothed on a 3D mesh contained in a box of 4x4 Mpc centered in the cluster. The grid size is 20kpc, which is much larger than the gravitational softening. This value ensures a sufficiently strong signal even for the ICL in the faint outer regions of the clusters (see below Section \ref{subsec:iraf}). We neglect dust reprocessing, since at z=0 our clusters are predicted to contain little dust, particularly in the ICL region, in reasonable agreement with observations \citep[see][]{gjergo2018}. 

In an analogous way we construct DM projected density maps by using the same grid definition. We compute for each MC three surface brightness maps and three DM density maps, using the projected properties along the main Cartesian axes of the simulation. 

\subsubsection{Fitting procedure and shape definition}
\label{subsec:iraf}

We derive what we call {\it isocontour shapes} of the simulated BCGs and clusters from the surface brightness and DM density maps described in the previous subsection. The {\it isocontour shapes} are obtained according to the fitted ellipses to the isophotal (isodensity) contours of the light (DM) distribution. The adopted fitting procedure is the {\footnotesize ELLIPSE} routine \citep{jedrzejewski1987} of the Space Telescope Science Analysis System (STSDAS), which Fourier analyses contours as a function of the azimuthal angle. Before we apply the  {\footnotesize ELLIPSE} routine, we mask the main subhalos which can hamper the fitting procedure. Only the central part of the satellite galaxies are masked with a typical size of $10\times10$ pixels.
 {\footnotesize ELLIPSE} output consist on a table that contains the fitted isocontour parameters: SMA, luminosity or density, centre coordinates, ellipticity\footnote{For this routine the ellipticity is defined as $1-q$, where $q$ is the usual defined major to minor semi-axis ratio.}, position angle of the SMA and their respective errors.

We show in Fig. \ref{fig:histrmax} the distributions of  $R_\text{MAX}$, which is defined as the maximum SMA up to which {\footnotesize ELLIPSE} is able to perform the fit. For the ICL, $R_\text{MAX}$ varies from $0.2R_{1000}$ up to $2R_{1000}$ with a median value of $0.75R_{1000}$. The DM surface density is fitted in general up to more than $1.0R_\text{MAX}$  since this tracer is dominant and extends to higher radii.

{\it Isocontour} shapes are characterised by $q$ and PA, obtained by averaging the parameters fitted by {\footnotesize ELLIPSE} up to $0.1R_{500}$ and $R_\text{LIM}$, related to the BCG and ICM regions, respectively. Here $R_\text{LIM}$ is the limting radius considered and satisfy $R_\text{LIM} \leq R_\text{MAX}$. The ICL shape in particular is characterised by the \textit{isocountour} parameters obtained from the fitted surface brighteness maps within the ICM. A similar approach of averaging PA of fitted ellipses to characterise the orientation of the ICL was used in  \citet{Kluge2021}. 

\subsection{Comparison between isocontour and inertial shape parameters}
\label{subsec:compineiso}

Fig. \ref{fig:bcgmaps} shows the BCG (left) and ICL (middle) surface brightness maps and the  DM surface density (right) for a late- and an early-formed cluster (upper and lower panel, respectively), along with the fitted ellipses corresponding to the inertial (solid line) and isocontour (dashed line) shape parameters. There is a general good agreement between both estimates of ellipticities and position angles. However, some differences between \textit{isocontour} and \textit{inertial} parameters can be appreciated, specially for the brightness distribution of the early-formed cluster. It is straightforward that the presence of substructure can affect the shape determinations.

\begin{figure}
\centering
    \includegraphics[scale=0.6]{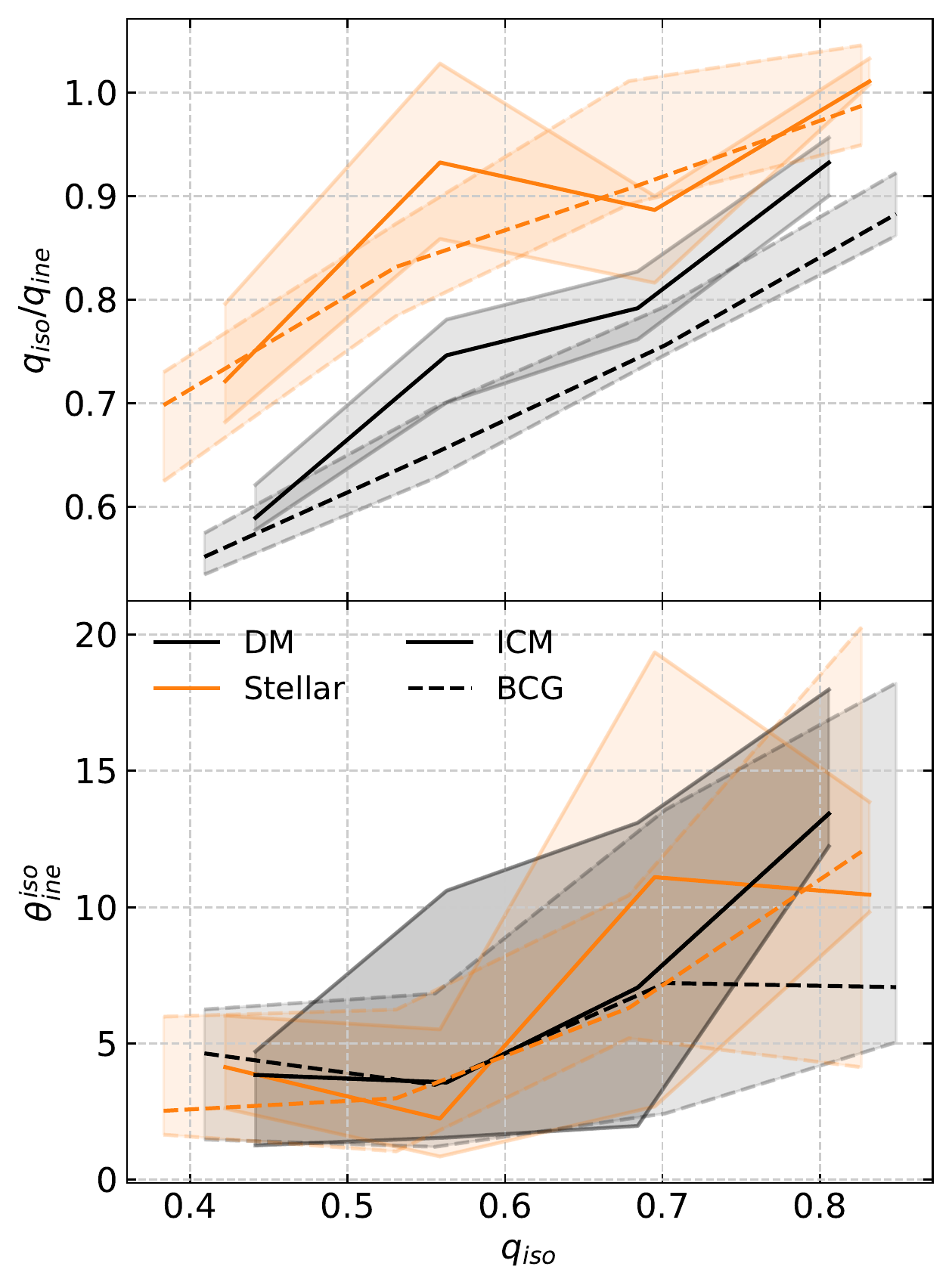}
    \caption{Ratio (upper panel) and misalignment (bottom panel) between the fitted \textit{inertial} and \textit{isocontour} parameters according to the \textit{isocontour} semi-axis ratio, $q_\text{iso}$. Orange and black lines correspond to the median values in $q_\text{iso}$ bins for the stellar and DM distributions, respectively. Solid and dashed lines correspond to the parameters constrained in the BCG and ICM regions, respectively. Shadow region enclose from 25th to 75th percentiles.
     }
    \label{fig:compare_methods}
\end{figure}

In order to study the differences in the shape parameters related to the methodologies applied to derive these quantities, we compare fitted \textit{isocontour} parameters derived in the BCG and ICM regions with the \textit{inertial} quantities computed using the star and DM particle positions within $0.1R_{500}$ and $R_{1000}$. To avoid biases introduced for constraining the shapes within different radial regions, we set $R_\text{LIM} = R_{1000}$ and for the stellar distribution we only consider the clusters with $R_\text{MAX} > 0.7R_{1000}$. In Fig. \ref{fig:compare_methods} we compare the projected semi-axis ratios derived according to the fitted \textit{isocontours} ($q_\text{iso}$) and \textit{inertial} tensor ($q_\text{ine}$) and the misalignment between the SMA orientations derived from both methodologies. In general, $inertial$ parameters tend to predict rounder shapes. There is a clear relation between the constrained elongation according the \textit{isocontours} and $q_\text{iso}/q_\text{ine}$. When \textit{isocontours} are rounder, semi axis-ratios tend to be more similar. This trend depend on the considered tracer and the cluster-centric region in the case of the DM. Rounder shapes also are related to higher misalignment angles since in this case the orientation is worse constrained. 

\section{Using observational tracers to asses the DM distribution}
\label{sec:results}

This section analyses how well the projected semi-axis ratio and orientation of the total dark matter distribution can be estimated from the galaxy members and the ICL. We also study the dilution of the lensing signal, introduced by the misalignment between the dark matter distribution and the tracers considered to estimate the orientation angle for weak-lensing stacking analysis. With this in mind, we aim to provide information regarding how well the usual proxies considered to align the clusters trace the dark matter distribution to minimize the introduced dilution.

\subsection{ICL as halo DM tracer}
\label{subsec:icl}
In the last years there have been a few works devoted to study, both from an observational and theoretical point of view, the capability of the ICL to trace the cluster dark matter density distribution  \citep[][]{,montestrujillo2019,Alonso-Asensio2020,Sampaio-Santos2021}. In this section we use the simulated clusters to assess this issue.

ICL shape parameters are obtained by fitting the isophotes of the surface brightness as detailed in \ref{subsec:iclmod} up to $R_\text{LIM} = R_\text{MAX}$. The following analysis compares the ICL semi-axis ratio and position angle with the {\it isocontour} parameters obtained for the DM within the same cluster-centric region.

\begin{figure}
\centering
    \includegraphics[scale=0.6]{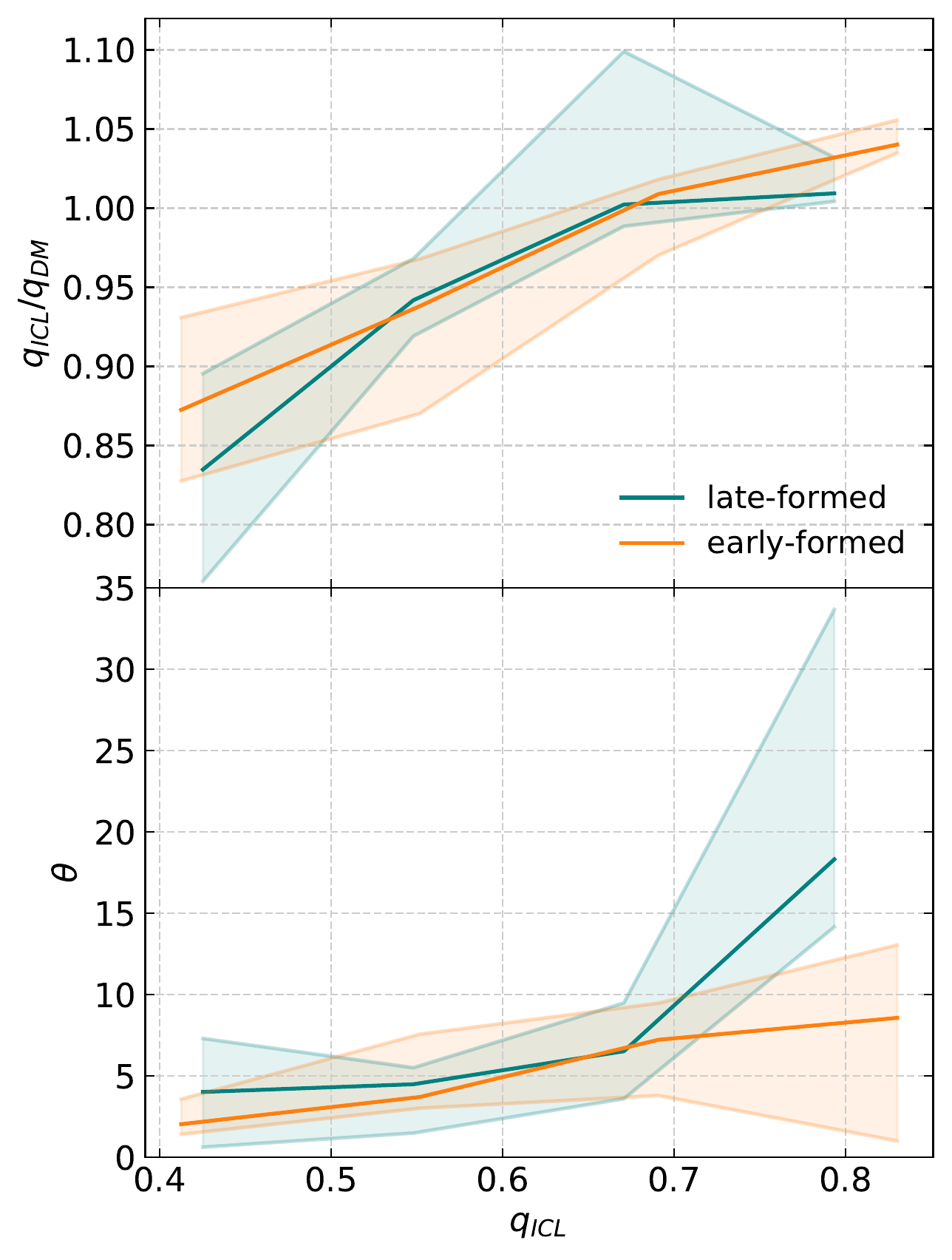}
    \caption{ICL and projected DM distributions median shape and misalignment angle as a function of the formation time for the MCs. Shadow regions in all panels enclose from 25th to 75th percentiles and inertial shapes are derived from particles within $R_{1000}$. 
    \textit{Upper panel:} projected semi-axis ratios $q$ derived for the DM, stars and the ICL. Solid lines stand for isocontour shapes of the ICL (red) and the DM (black), while dashed lines correspond to inertial shapes measured with stars (orange) and DM particles (grey) within $R_{1000}$. Inertial shapes are systematically more rounded than isocontour ones. 
    \textit{Bottom panel:} Median misalignment angles between the ICL and the star particle distribution (orange dashed line), and between the ICL and the isocontour (inertial) derivation of the DM shape  in solid black (dashed grey) lines.  
     }
    \label{fig:ICL}
\end{figure}

In Fig. \ref{fig:ICL} we compare the surface brightness and DM density shape parameters for the MCs. We split the cluster sample according to the median formation time (4.9Gyr) as later- and early-formed clusters. The stellar component is in general more elongated than the DM distribution for all the clusters, in agreement with the differences in the shapes observed between the stellar and the DM particles distributions (see Appendix \ref{A:stars}). Also, discrepancies between the ICL and DM elongation increase as the ICL is more elongated. This trend is observed for later- and early-formed cluster as well.  

According to the misalignment angle, the DM SMA position angle is well constrained by the ICL, with a median misalignment lower than $10^\circ$ for early-formed clusters. The median as well as the dispersion increase for rounder ICL shapes, since the position angle is worse constrained for more spherical distributions. 

\begin{figure*}
    \includegraphics[scale=0.6]{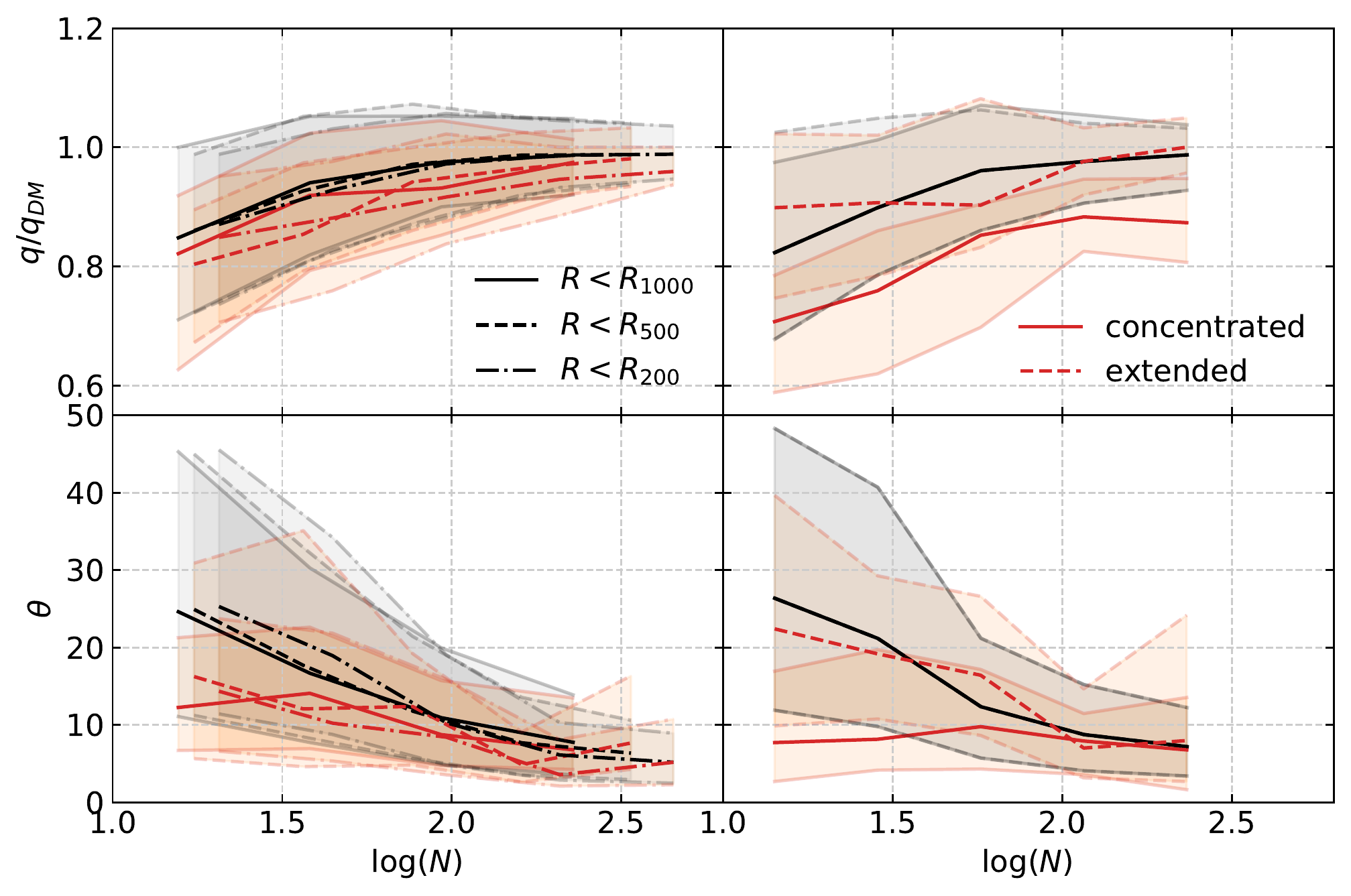}
    \caption{Left panels: Red (black) lines correspond to median $q$ ratio and misalignment angle between the galaxy member distribution (randomly selected DM particles) and the total DM particle distribution in bins of $\log(N)$, where $N$ is the number of subhalos (particles) within $R_{1000}$, $R_{500}$  and $R_{200}$.  Shaded regions enclose from 25th to 75th percentiles. Right panels: Same as in the left panels but taking the subhalos and particles within $R_{200}$ and considering the subhalos classified as \textit{concentrated} and \textit{extended}.
     }
    \label{fig:shape_gx_dm_2D}
\end{figure*}

Taking into account these results, using the ICL to derive the cluster DM shape would be biased to higher elongations.  On the other hand, the ICL shape is well orientated with the DM, mainly when considering early-formed clusters or for more elongated ICL distributions. For rounder late-formed clusters, the presence of substructure could be affecting the shape fitting (Fig. \ref{fig:bcgmaps}). It is important to highlight that using the ICL shapes as an observational test of DM cluster shapes needs to take into account the methodology applied in order to constrain shape parameters, since as we show in \ref{subsec:compineiso} differences between the usual inertial parameters are expected with respected to the fitted isocontours.

\subsection{The galaxy distribution as halo DM tracer}
\label{subsec:gx}

Here we discuss how the galaxy member distribution can constrain the total DM cluster halo shape. We consider the \textit{inertial} shape parameters obtained using the galaxy members and the DM particle positions for the analysis. Thus, we test how well the dark matter distribution is traced by the total cluster members and the two subsamples, \textit{extended} and \textit{concentrated} galaxies, defined in \ref{subsec:members}.

One of the main caveats of constraining the DM distribution ellipticity according to the galaxy member positions is the noise bias,  introduced from measuring the shape parameters with a low finite number of members. This bias tends to boost the estimated elongation.
Measured projected ellipticities can be corrected by using simulations, as proposed by \citet{wang2018}. Nevertheless, correcting for this effect is not straightforward since there is a co-dependence between the number of identified members, the unbiased distribution and the introduced noise. The noise bias is higher when considering fewer galaxies and when the underlying distribution is rounder. Therefore, some considerations regarding the underlying matter distribution have to be assumed to quantify this bias. To account for this effect, we compute the shape parameters of the DM using only $N$ randomly selected particles, where $N$ is the number of galaxies considered to derive the shape parameters of the galaxy distribution. We make 1000 realisations, and then, for each cluster, we obtain the median value of the shape parameters computed from the random realisations.

Another bias introduced by this approach is the inclusion of interlopers in the galaxy member sample, i.e. wrongly identified members unrelated to the cluster halo. This bias lowers the estimated ellipticity value and increases the misalignment. We consider this effect by taking into account the identified galaxies within a cylinder limited by the re-simulated region ($\sim 40$kpc), as detailed in \ref{subsec:members}.

In the upper left panel of Fig. \ref{fig:shape_gx_dm_2D} we show, within different radii, the ratio between the randomly selected DM particles (black) or galaxies (red) and the total DM distribution $q_{DM}$, binned according to the number of galaxies considered in each cluster, $N$. As expected, the introduced discreetness noise, biases the projected semi-axis ratio to lower values. This effect significantly decreases when considering more than $\sim$ 100 galaxies, in fact the median ratio becomes higher than 0.9. 
The fact that red curves are systematically bellow the black ones, regardless of the number of galaxies considered, indicates that the galaxy distribution is actually more elongated that the DM distribution. 
In the bottom left panel we show the median angle between member galaxies and DM (red) and between the randomly selected DM particles and the total DM distribution (black). The later gives us an estimation of the impact of the noise bias in the determination of the DM halo PA. As expected, the smaller the number of selected particles $N$, the larger the $\theta$ angle and hence the stronger the noise bias. If we now concentrate in the red curves, we find that no matter the number of galaxies $N$, the galaxy distribution is well aligned with the DM halo. Moreover it is even better aligned than the random DM particles themselves, which could be related to the fact that the random DM particles are selected from a more rounder distribution than that of the galaxies, as shown before.
Finally, the right-hand panels of Fig. \ref{fig:shape_gx_dm_2D} show that \textit{extended} galaxies' distribution is more spherical than that of more \textit{concentrated} galaxies, and that the latter features a better alignment with the SMA of the total DM. 

The observed shape trends of the galaxy distributions discussed in this subsection, show a similar behaviour in the full 3D analysis presented in Appendix \ref{A:gxs}. 
In this case the galaxy member sample does not include interlopers and the differences in the shape parameters between the randomly selected DM particles and the galaxy distribution are more significant. Considering the 3D sphericity, the galaxy members follow a more elliptical distribution and are well-aligned with the DM SMA. Also, \textit{concentrated} galaxies follow an even less spherical and more prolate distribution than the total DM and the \textit{extended} galaxies. 

We notice that the observed differences found for \textit{concentrated} and \textit{extended} galaxies could be related to the cluster accretion time of these two subhalos sets. Tidal effects during the accretion of subhalos may efficiently induce removal of material from their outskirts making
them more compact \citep{Diemand2007,Kuhlen2008,Springel2008}. Therefore, \textit{concentrated} galaxies are expected to be accreted earlier compared to the \textit{extended} ones. The observed elongated distribution for the more \textit{concentrated} galaxies and its good alignment with the DM halo SMA are also in agreement with observational analysis. 
In fact, these studies provide evidence that red satellites show a more anisotropic distribution and are more tightly aligned with the central galaxy of the host halo \citep{yang2006,Wang2010} than the blue counterparts. It can be argued that this alignment is related to evolution effects after accretion \citep{Wang2014}. Moreover, \citet{Gonzalez2021} reported higher elongation values for clusters using weak lensing stacking techniques, when galaxies with a high membership probability are used to align the systems. This population of galaxies tends to be redder and has lower cluster-centric distances, in agreement with our findings for \textit{concentrated} galaxies.

According to our results, projected semi-axis ratio estimates of the dark matter halo derived from the galaxy member distribution could be biased to lower values, even if the values are corrected considering the noise bias and interlopers are included in the samples. Differences could be even larger if we consider early-type galaxies and low-mass poor clusters. Therefore, derived shapes can be affected by the particular characteristics of the identified members. Thus, the parameters can depend on the identification algorithm applied. On the other hand, for galaxy clusters with less than 100 identified galaxy members, the orientation angle of the dark matter distribution could be better constrained, within a median value $\lesssim 10^\circ$, by using early-type galaxy positions.

\subsection{Dilution Factor}

\begin{figure*}
    \includegraphics[scale=0.6]{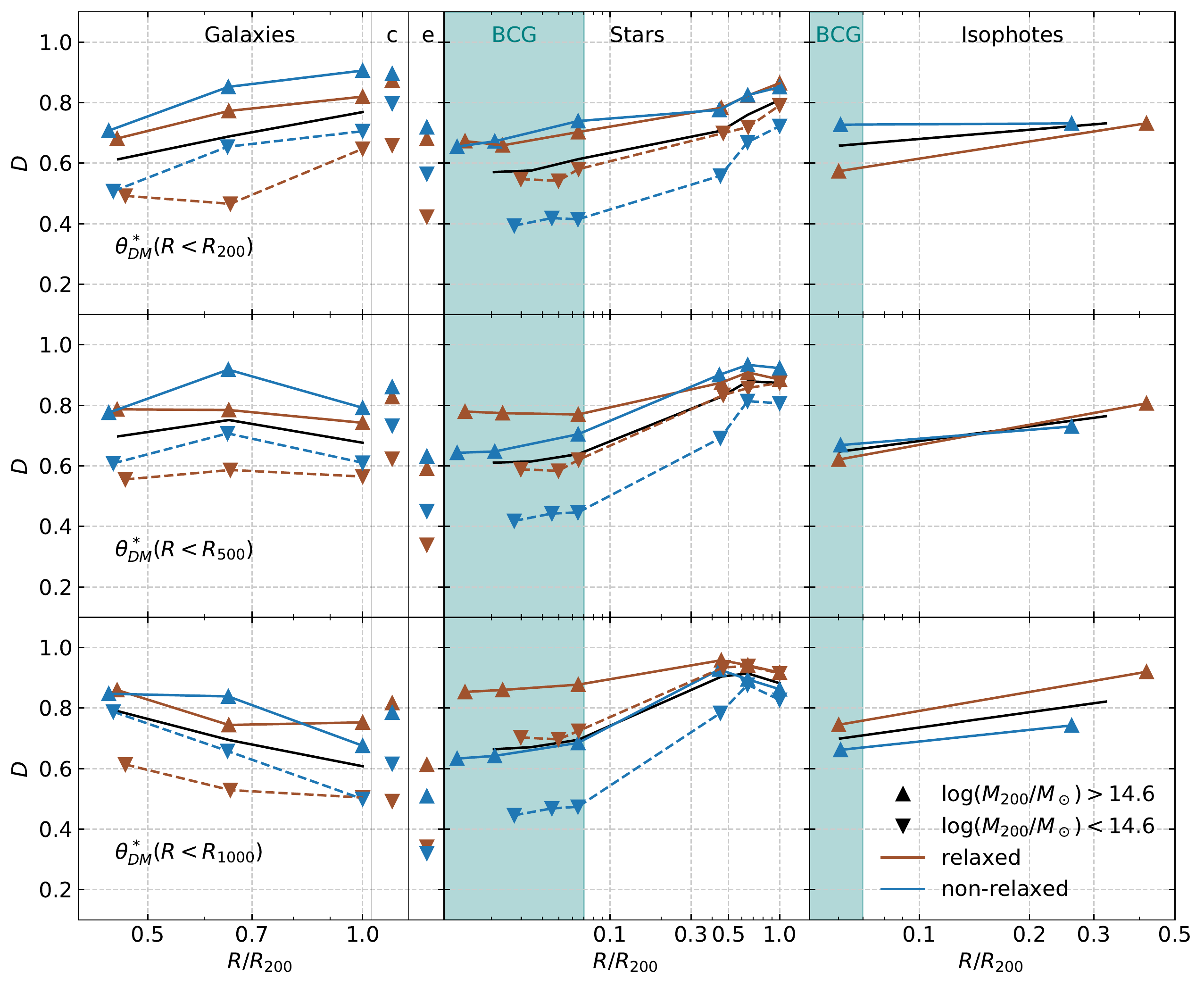}
    \caption{Dilution factor, $D$, computed according to average of $\cos(2 \Delta \theta)$, where $\Delta \theta$ is the projected angle between the galaxy (left column) and stellar (right column) distributions within $R/R_{200}$ and the dark matter particle distribution within $R_{200}$, $R_{500}$ and $R_{1000}$ (upper, middle and bottom panels, respectively). $D$ is computed for the total sample of clusters (solid black line), high-mass (solid color line) and low-mass clusters (dashed color line) and relaxed (red lines) and non-relaxed clusters, classified according to the relaxation criteria $M_\text{sat}/M_\text{BCG}$ (circles), $D_\text{offset}$ (squares)  and $\Delta V$ (triangles). }
    \label{fig:D_R}
\end{figure*}

As shown in the previous subsections, constraining the DM cluster shape component from the population of galaxy members and the ICL could result in biased estimates since these luminous tracers follow a more elongated distribution in comparison to the DM. The use of gravitational lensing techniques can be a useful approach to constrain cluster halo shapes given that they are sensitive to the total mass distribution, dominated by the DM component. Nevertheless, lensing shape estimates could also be biased. In particular, projected cluster ellipticity measurements using stacking weak-lensing techniques are affected by the misalignment $\theta$ of each cluster, defined as the angle between the orientation of the DM distribution surface density and the estimated angle used to align and combine the clusters. Therefore, the lensing signal is diluted by a factor given by the mean of $\cos(2\theta)$ for the sample of clusters considered \citep{Clampitt2016,vanUitert2017}.  This results in measured ellipticities biased to lower values compared to the projected dark matter distribution:
\begin{equation}
    \epsilon_\text{WL} = \langle \cos(2 \theta) \rangle \times \epsilon_\text{DM} = D \times \epsilon_\text{DM},
\end{equation}
where $\epsilon_\text{DM}$ is the projected ellipticity of the total dark matter distribution and $D$ is the dilution factor. Since the true misalignment for each cluster is unknown in observational studies, $D$ can be estimated by assuming a Gaussian distribution for the misalignments with a given dispersion computed according to mock realisations. 

Here we inspect the actual mean dilution factor computed for different ensembles of clusters selected according to the mass and the relaxation state indicator to provide information regarding which tracer may optimise the orientation estimate, thus boosting the $D$ value. For this purpose, we compute the misalignment between the estimated orientation of different tracers and the inertial orientation of the DM particle distribution for the study. To this end
we use the inertial DM shape estimates since these parameters are obtained for the whole sample of clusters and up to larger cluster-centric regions, as compared to the determination based on the isodensity contours. The spacial range is particularly important, as the inertial determinations cover a range
comparable to those used in the weak-lensing analysis. 
Moreover, the lensing observables are affected by the whole DM particle distribution and then affected by the cluster substructure and subhalos, which are masked in the isocontour approach.
Nevertheless, we explore the differences between considering the DM {\it isocontour} parameters instead of the {\it inertial} ones up to $R_{1000}$, by computing the dilution values for the MCs using the different tracers. As expected, we obtain larger differences mainly for non-relaxed clusters, showing higher values of $D$ when using the {\it isocontour} DM shapes. Nonetheless, the general trends are preserved. 

In Fig. \ref{fig:D_R} we show the dilution factor computed using the misalignment angles between the total DM particle distribution and the considered tracers: cluster galaxies (left), star particles (middle) and surface brightness distributions (right). We show the results for different samples of clusters: relaxed and non-relaxed, classified according to the cluster relaxation proxy described in \ref{subsec:relax}; and high- and low-mass clusters, classified taking into account the median value of the mass distribution. In the case of the ICL, since shape estimates were derived only for MCs, we show results only for high-mass clusters. DM inertial shapes, as well as the tracer shape parameters, are compared within different cluster-centric distances. In general, for high-mass clusters the dilution factor is higher than for the low-mass sample. Also, this factor is higher when the adopted tracer distribution extends to the same radius as for the dark matter particles.

In low-mass non-relaxed cluster, galaxy members better follow the orientation of the DM particles than for low-mass relaxed ones. A similar trend is present among high-mass clusters. We also show the obtained dilution values when considering \textit{concentrated} and \textit{extended} galaxies as tracers within $R_{200}$. As expected from the results in \ref{subsec:gx}, higher dilution values are obtained when considering \textit{concentrated} galaxies as tracers, due to a better alignment between these galaxies and the DM SMA.

In contrast to the galaxy member distribution, the stellar distribution is in general better aligned with the total DM particles when considering relaxed halos instead of non-relaxed ones, especially within the BCG region. As we move towards the cluster centre (from the upper panels to the bottom in Fig. \ref{fig:D_R}), those classified as relaxed show that the stellar distribution aligns better with the dark matter, especially in the BCG region. The same trend is present when considering the fitted isophotes of the surface brightness density. 

Weak lensing studies are mainly sensitive to the surface mass distribution up to $R_{500}$ and $R_{200}$, depending on the limiting projected distance at which the lensing surface density contrast profiles are fitted \citep{Harvey2021}. According to our results, among the considered tracers, the galaxy distribution is the most suitable proxy to constrain the orientation of the dark matter distribution. Also, it is more efficient when the galaxies are restricted at the same radius as for the DM is constrained. Although this result seems straightforward, restricting the galaxy distribution to a smaller radius results in fewer tracers. This reduction increases the sampling noise, but this is compensated by a better alignment between the galaxies and the dark matter distributions. For low mass clusters, the dilution can be significant with $D \lesssim 0.6$. For these clusters, a substantial improvement of the dilution can be achieved when considering the stellar distribution up to the outer radius. Nevertheless, fitting the ICL for these low-mass systems is very challenging given that longer exposure times are required, since the surface brightness of the diffuse light shows an increasing dependence on cluster total mass \citep{Sampaio-Santos2021}. The BCG orientation also provides moderate dilution values, specially when considering relaxed clusters. Indeed, previous studies have found that the BCG is well aligned with the cluster DM halo from at least 8 Gyr and that this alignment is tighter than with the galaxy member distribution \citep[][]{Ragone2020}. Another alternative to improve the dilution for these clusters could be using only \textit{concentrated} galaxies to estimate the SMA orientation. 

\section{Summary and Conclusions}
\label{sec:conclusion}

In this work, we studied the galaxy cluster shapes using cosmological hydro-simulations including sub-grid baryon physics. We considered the distributions of different cluster components, stars, galaxies and dark-matter. Moreover, we adopted two different approaches to compute the shape parameters: fitted projected isocontours and inertial parameters obtained from the position of the tracers using the inertial tensor. We also constrained the shape distribution up to a set of radial distances, thus obtaining information from different cluster-centric regions.

We explored how the shape and orientation of the DM cluster halo can be constrained using the observable tracers, as the galaxy member positions and the ICL. According to our results, cluster galaxies tend to follow a less-spherical distribution compared to the DM particles at all the considered radius (from $R_{1000}$ up to $R_{200}$). They are also well aligned with the SMA of the DM distribution with a median misalignment of $\sim 10^\circ$, including those clusters with a low number of members despite the introduced noise-bias. We also inspected the differences between the selected samples of galaxy members, considering \textit{concentrated} and \textit{extended} galaxies, classified according to the DM subhalo distribution. We obtain that \textit{concentrated} galaxies follow a more elongated distribution than the dark matter, with a median projected semi-axis ratio of $\lesssim 0.85$ times the semi-axis ratio of the projected DM particle distribution. For lower mass galaxy clusters (those with less than 100 identified galaxy members), \textit{concentrated} galaxies also show a significantly better alignment with the DM orientation than the obtained for the distribution of \textit{extended} galaxies. 

Regarding using the ICL to constrain the DM distribution, we compared the projected semi-axis ratio from the surface brightness density fitted within the ICM with the elongation derived from the projected DM isodensity and the projected particle distribution. The ICL semi-axis ratio is in median bias to lower values, indicating that the brightness distribution is, in general, more elongated. This result is in correspondence with the observed differences between the stellar and DM distribution. We also obtain a well-alignment between the ICL and the surface DM density, with median misalignment angles of $\sim 5^\circ$. Higher misalignment values arise when the ICL shape distribution is rounder, especially for clusters with low formation times. 

Finally, we computed the dilution factor introduced when deriving the projected ellipticity of the surface density distribution using weak-lensing stacking techniques. This factor lowers the estimated elongation due to the misalignment between the estimated and the real main surface density orientation. We derived the dilution values for different subsamples of clusters, obtained from the misalignment between the total DM particle distribution and the tracers up to several cluster-centric regions. Using the galaxy member positions to estimate the DM orientation offers a suitable approach to perform the stacking.

Cluster shapes are a key property of these systems that provide very useful information on their formation history and may serve as cosmological tests. Based on cosmological simulations, this work presents a detailed analysis of the interplay between dark matter, intracluster light, and galaxy distributions in providing cluster shapes. Our main motivation was aimed to predict the way observational tracers follow the total dark-matter distribution. The results following our analysis provide valuable information for observational studies aimed at constraining the shapes of galaxy clusters.

\section*{Acknowledgements}
We warmly thank Matthias Kluge for the helpful feedback and useful discussion.
This project has received funding from the Consejo Nacional de Investigaciones Cient\'ificas y T\'ecnicas de la Rep\'ublica Argentina (CONICET), from the Secretar\'ia de Ciencia y T\'ecnica de la Universidad Nacional de C\'ordoba - Argentina (SECyT-UNC), and from the European Union's Horizon 2020 Research and Innovation Programme under the Marie Sklodowska-Curie grant agreement No 734374. MM is partially supported by FAPERJ and CNPq Fora Bozo.
Simulations have been carried out in the Centro de Computaci\'on  de alto desempe\~no (CCAD) de la UNC, which is part of the Sistema Nacional de Computaci\'on de Alto Desempe\~no del Ministerio de Ciencia, Tecnolog\'ia e Innovaci\'on (SNCAD-MinCyT, Argentina), and at the computing centre of the Istituto Nazionale di Astrofisica (INAF-Italia). We acknowledge the computing centre of INAF-Osservatorio Astronomico di Trieste, under the coordination of the Calcolo HTC in INAF - Progetto Pilota (CHIPP) \citep{bertocco2019,taffoni2020}, for the availability of computing resources and support.

\section*{Data Availability}
The data underlying this article will be shared on reasonable request to the corresponding author.




\bibliographystyle{mnras}
\bibliography{bib} 




\appendix

\section{Inertial shape parameters}
\label{A:shape}

Inertial shape parameters are computed according to the semi-axes derived from the particle distribution inertia tensor:
\begin{equation} \label{eq:tensor}
\mathcal{T}_{ij} = \frac{1}{N} \sum_n x_{n,i} x_{n,j},    
\end{equation}
where $x_{n,i}$ and $x_{n,j}$ are the $i^{th}$ and $j^{th}$ component of the $n^{th}$ particle position vector relative to the halo centre and $N$ is the number of particles. For the 3D particle distribution $i,j = 1,2,3$  and the corresponding 
values of the semi-axes ($a > b > c$) are the square roots of the eigenvalues of the inertia tensor, while their directions ($\hat{a}, \hat{b}, \hat{c}$) are defined by the normalised eigenvectors. Likewise, for the 2D particle distribution $i,j = 1,2$, we obtain the moduli ($a^* > b^*$) and directions ($\hat{a}^*, \hat{b}^*$) of the semi-axes from
the eigenvalues and the eigenvectors of the inertia tensor, respectively.

From the semi-axes we compute the shape parameters associated 
to the 3D and 2D particle distribution. We define the triaxiality and sphericity parameters from 
the 3D density distribution as:
\begin{equation}
    T = \frac{a^2 - b^2}{a^2 - c^2}\,\, \text{  and  }\,\,S = \frac{c}{a}.
\end{equation}
For spherical distributions $S = 1$ while $T$ will be undefined. Higher values of $T$ ($T$ close to 1) 
indicate a prolate distribution while lower values ($T$ close to 0) indicate an oblate distribution.
For the 2D shape distribution we compute the projected semi-axis ratio:
\begin{equation}
    q = b^*/a^*.
\end{equation}
which is analogous to the 3D sphericity parameter, i.e. for a spherical distribution $q=1$.

We also inspect the 3D alignment angle between the major semi-axes obtained from different tracers:
\begin{equation}
    \theta^{3D} = \arccos{(|\hat{a}_{t_1} \cdot \hat{a}_{t_2}|)},
\end{equation}
where $\hat{a}_{t_1}$ and $\hat{a}_{t_2}$ are the major semi-axis versors computed according to the distribution of the considered tracers $t_1$ and $t_2$ within a particular radius, respectively. In an analogous way, we define the 2D alignment angle $\theta$ considering the versors computed according to the 2D distribution, $\hat{a}^*_{t_1}$ and $\hat{a}^*_{t_2}$. If the 3D semi-axes were randomly oriented
we expect a median $\theta = 60^\circ$ and 25 per cent–75 per cent percentiles of $\sim 41.4^\circ$ $\sim 75.5^\circ$ , respectively. On the other hand, for a uniform random distribution of projected orientation we expect a median $\theta^* = 45^\circ$ and 25 per cent–75 per cent percentiles of $\sim 22.5^\circ$ $\sim 67.5^\circ$, respectively. The parameters defined above are derived considering the position vectors of the galaxies, stars and the dark matter particles as tracers (see Sec. \ref{sec:params}).

\section{Subhalo effect in the shape of the dark matter distribution}
\label{A:HvsDM}

\begin{figure}
\centering
    \includegraphics[scale=0.6]{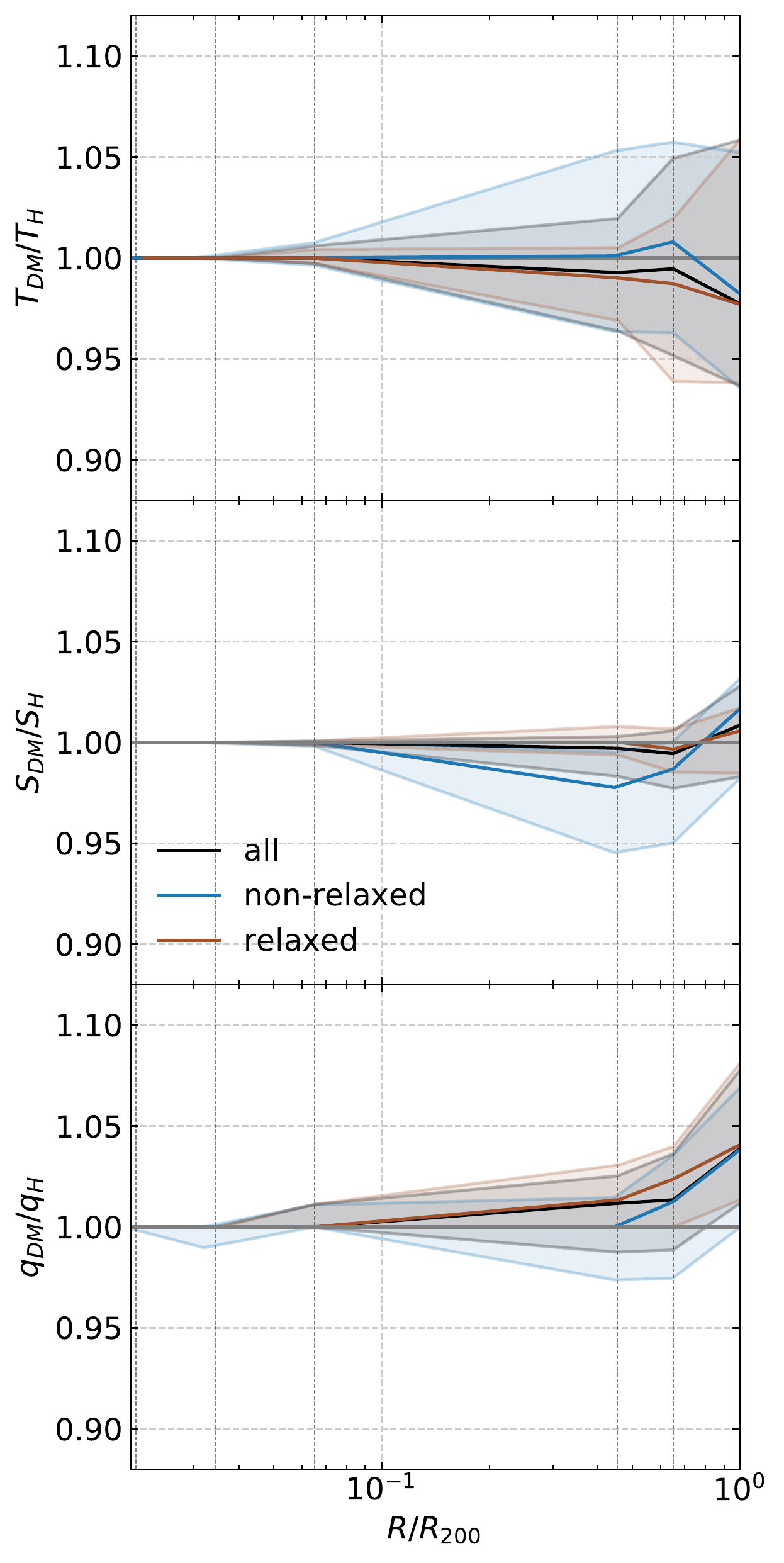}
    \caption{Ratio of the shape parameters, 
    triaxiality (upper panel), sphericity (middle panel) and projected semi-axis ratio (bottom panel), between the total dark matter and the halo particle distributions, computed considering the particles within the 
    radius shown in the x-axis. Black solid lines correspond to the median values of the ratios for the whole sample of clusters. Blue and red lines correspond to non-relaxed and relaxed halos, respectively. Shaded regions enclose from 25 to 75 per cent percentiles. The grey solid line set at one is plotted as reference.}
    \label{fig:DMvsH}
\end{figure}

Here we inspect the differences between the shape parameters derived from the total dark matter particle distribution (DM) and the distribution of the halo dark matter particles (H, i.e. total dark matter particle distribution without the particles that belong to the identified subhalos). 
According to the median triaxiality ratio between the total and the halo dark matter particles within all the radii considered 
(upper panel of Fig. \ref{fig:DMvsH}), there is a good correspondence in this shape parameter computed for both distributions with an increasing dispersion at higher radii.
In the case of the sphericity, halos classified as non-relaxed show the largest differences between both distributions at $\sim R_{1000}$ ($\sim 0.45\,R_{200}$), such that
the halo particle distribution is slightly rounder than the total dark matter particle distribution (middle panel of Fig. \ref{fig:DMvsH}). On the other hand, projected shapes of the total dark matter distributions are rounder with increasing radius than the halo particle distribution. 

\begin{figure}
\centering
    \includegraphics[scale=0.6]{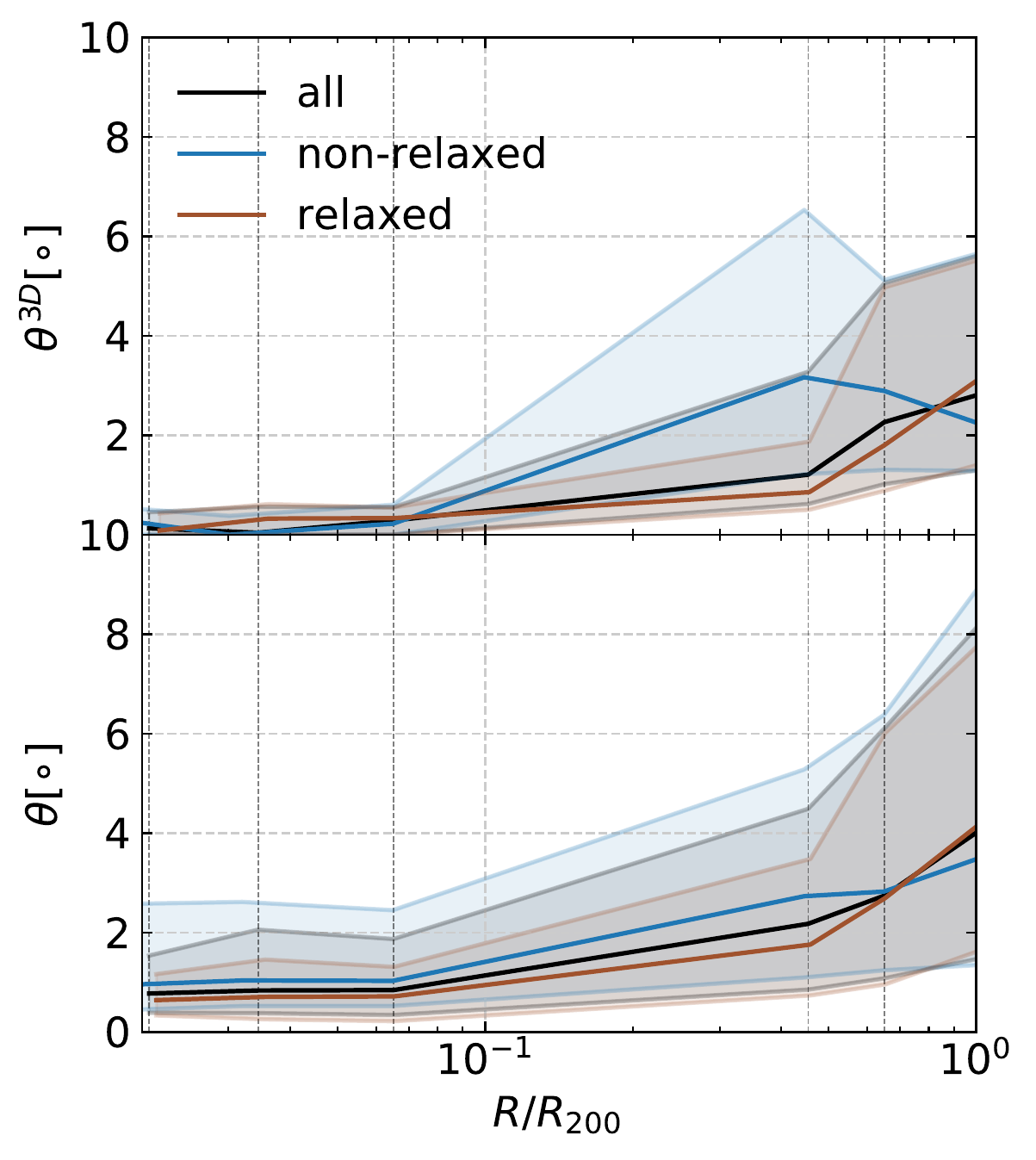}
    \caption{3D (upper panel) and 2D (lower panel) alignment angle between the major semi-axes of the total dark matter and the halo distributions, computed considering the particles within the specified radius in the x-axis. Black solid lines correspond to the median values of the computed angles for the whole sample of clusters. Blue and red lines correspond to non-relaxed and relaxed halos, respectively. 
    Shaded regions enclose from 25 
    to 75 per cent percentiles. The grey solid line set at one is plotted as reference.}
    \label{fig:DMvsH_theta}
\end{figure}

We also inspect the alignment angle between the total DM and the H particle distributions, for relaxed and non-relaxed
clusters. We notice a higher misalignment in 3D at $R_{1000}$ for non-relaxed halos (Fig. \ref{fig:DMvsH_theta}). 3D and 2D misalignment increase with radius, but median values for the whole range of considered radii are lower than $5^\circ$.

\section{Relation between dark matter shape parameters and halo properties}
\label{A:SP}

\begin{figure}
\centering
    \includegraphics[scale=0.6]{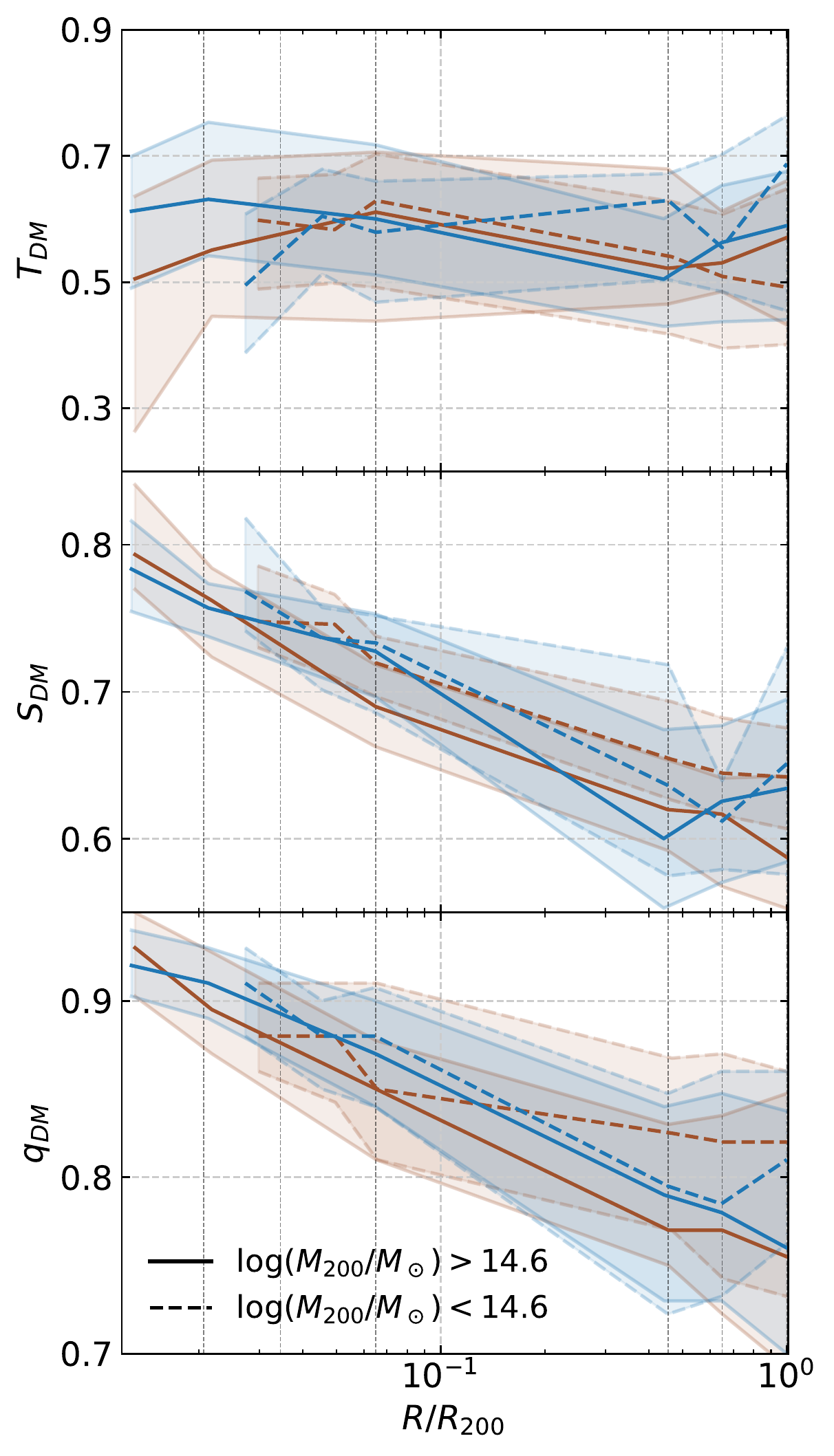}
    \caption{Median values of the shape parameters derived for the total dark matter particle distribution within the considered radius scaled with $R_{200}$. Shaded regions enclose from 25$^{\rm th}$ to 75$^{\rm th}$ percentiles. The total cluster sample is split according to the cluster mass, $M_{200}$, and relaxation state. Higher and lower mass cluster relations are plotted in solid and dashed lines, respectively. Relaxed (red lines) and non-relaxed (blue lines) classified according to $M_\text{sat}/M_\text{BCG}$, $D_\text{offset}$ and $\Delta V$ relations are shown in the left, middle and right panels, respectively. Vertical lines correspond to the median scaled {\bf radii} for the total sample of clusters listed in Table \ref{tab:radius}.}
    \label{fig:DM}
\end{figure}

In this section we inspect the relation between the shape parameters computed at different radii and the cluster $M_{200}$ masses. Since the scatter between the mass and the shape parameters is expected to be connected with the halo formation history, we also inspect how this relation depends on the halo relaxation. In Fig. \ref{fig:DM} we show the median values of the shape parameters from the dark matter particle distribution, measured at different limiting radii, for halo masses above and below $4 \times 10^{14} M_{\sun}$. 

\begin{figure}
\centering
    \includegraphics[scale=0.6]{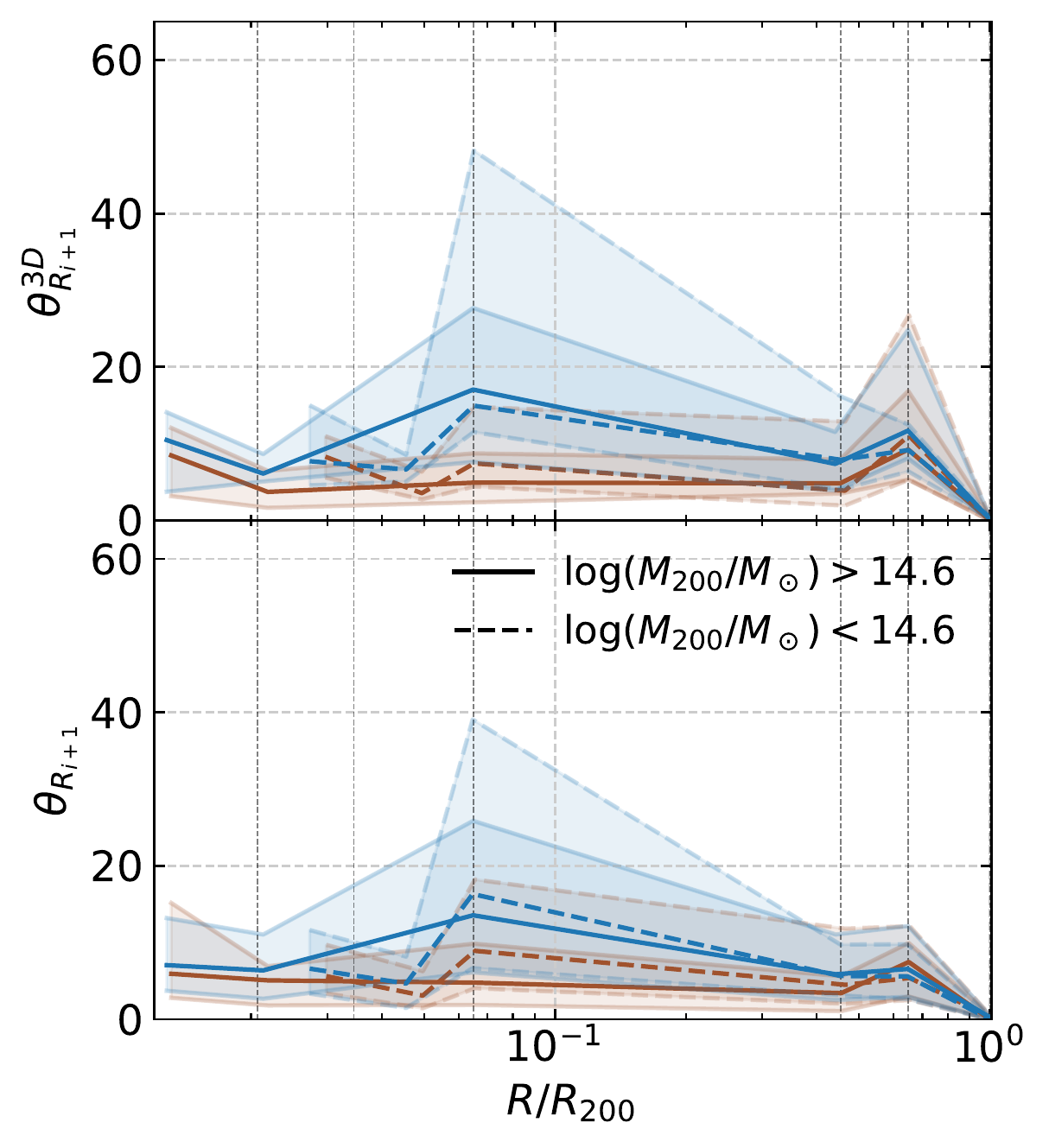}
    \caption{Median values of the 3D ($\theta^{3D}$) and 2D ($\theta$) misalignment angles between the DM major semi-axes computed within each radius and the subsequent radius. The total cluster sample is split according to the cluster mass, $M_{200}$, and relaxation state. Higher and lower mass cluster relations are plotted in solid and dashed lines, respectively. Relaxed (red lines) and non-relaxed (blue lines) classified according to $M_\text{sat}/M_\text{BCG}$, $D_\text{offset}$ and $\Delta V$ relations are shown in the left, middle and right panels, respectively. Vertical lines correspond to the median scaled radii for the total sample of clusters listed in Table \ref{tab:radius}.}
    \label{fig:tDM}
\end{figure}


There is a clear tendency for clusters to be rounder in the inner regions in 3D, as well as in the projected shapes. The opposite tendency is observed in dark matter only simulations \citep{Despali2017}. 
The inclusion of baryon physics causes the dark matter to become significantly rounder, specially in the inner regions \citep{Chua2019}. As noted
in previous studies, lower mass clusters are systematically rounder than high mass ones. 
The highest differences are seen 
at the outskirts in projected shapes. Differences between median sphericity values for relaxed and non-relaxed samples are not significant. 

Finally we inspect the 3D and projected misalignment angles  
between the dark matter SMA position angle computed 
in regions enclosed within different radii. In particular, we 
obtain the misalignment between the position angles 
computed within $R < R_{i}$ and $R < R_{i+1}$ with $R_{i} = 30\text{kpc}, 50\text{kpc}, 0.1R_{500}, R_{1000}, R_{500}, R_{200}$. We refer to these misalignement angles in 3D and 2D as $\theta^{3D}_{R_{i+1}}$ and $\theta_{R_{i+1}}$. The results are shown in Fig. \ref{fig:tDM}. Higher median values of $\theta^{3D}_{R_{i+1}}$ and $\theta_{R_{i+1}}$ occur for SMA computed considering the particles within $R < 0.1R_{500}$ and $R < R_{1000}$, specially for non-relaxed clusters, i.e. between the BCG region and the ICM. For higher mass and relaxed clusters, the median misalignment angles are lower for all the considered radii, indicating that for this sample there is a better alignment between the dark matter distribution from the BCG region up to $R_{500}$. For all the clusters considered there is also a rise in median angles at $R_{500}$.   

\begin{figure}
\centering
    \includegraphics[scale=0.6]{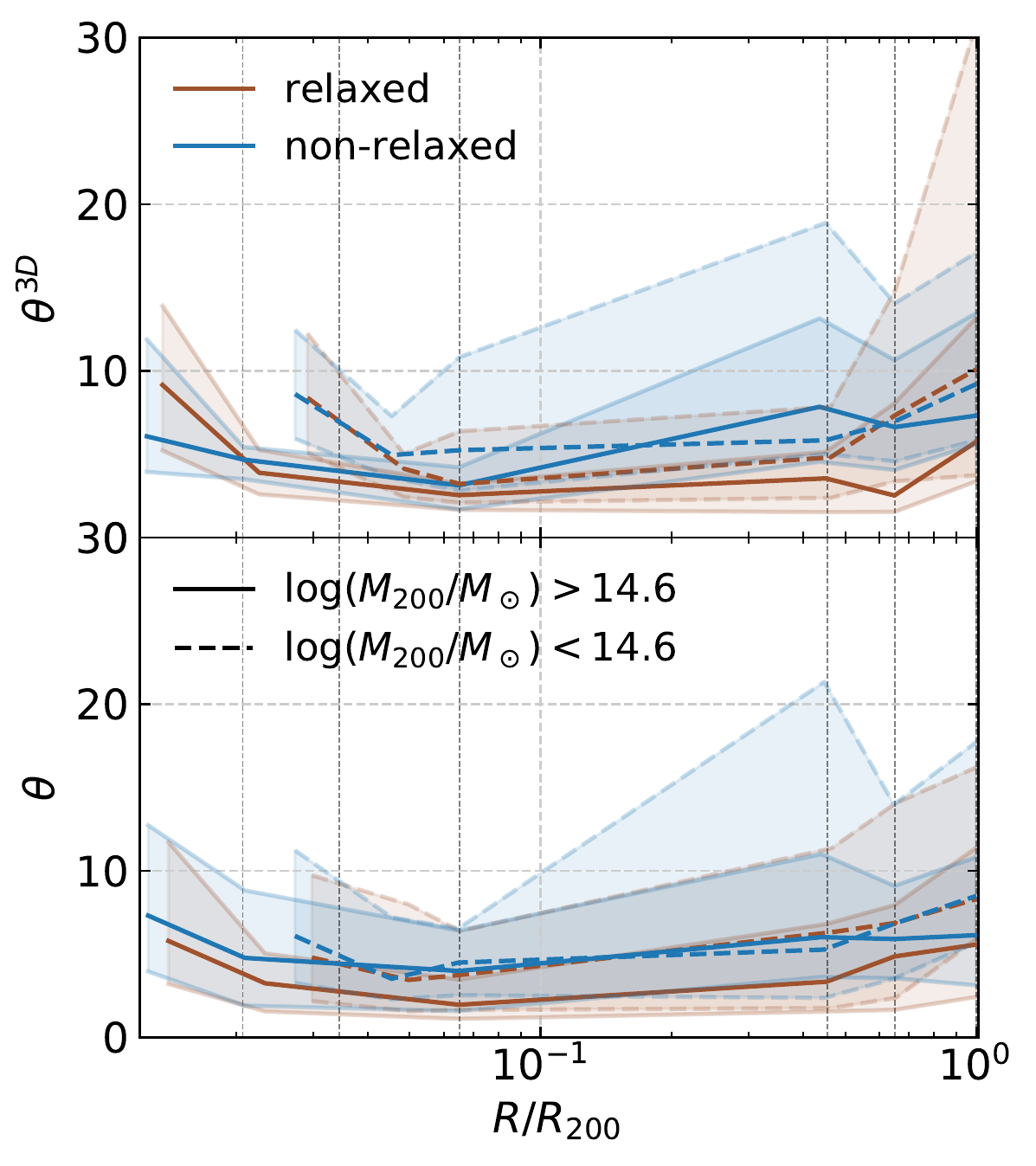}
    \caption{
    Same as in Fig. \ref{fig:tDM} but considering the 3D ($\theta^{3D}$) and 2D ($\theta$) misalignment angles of major semi-axes computed for the stellar and the total dark matter particle distributions.}
    \label{fig:tStarsDM}
\end{figure}

\begin{figure*}
\centering
    \includegraphics[scale=0.6]{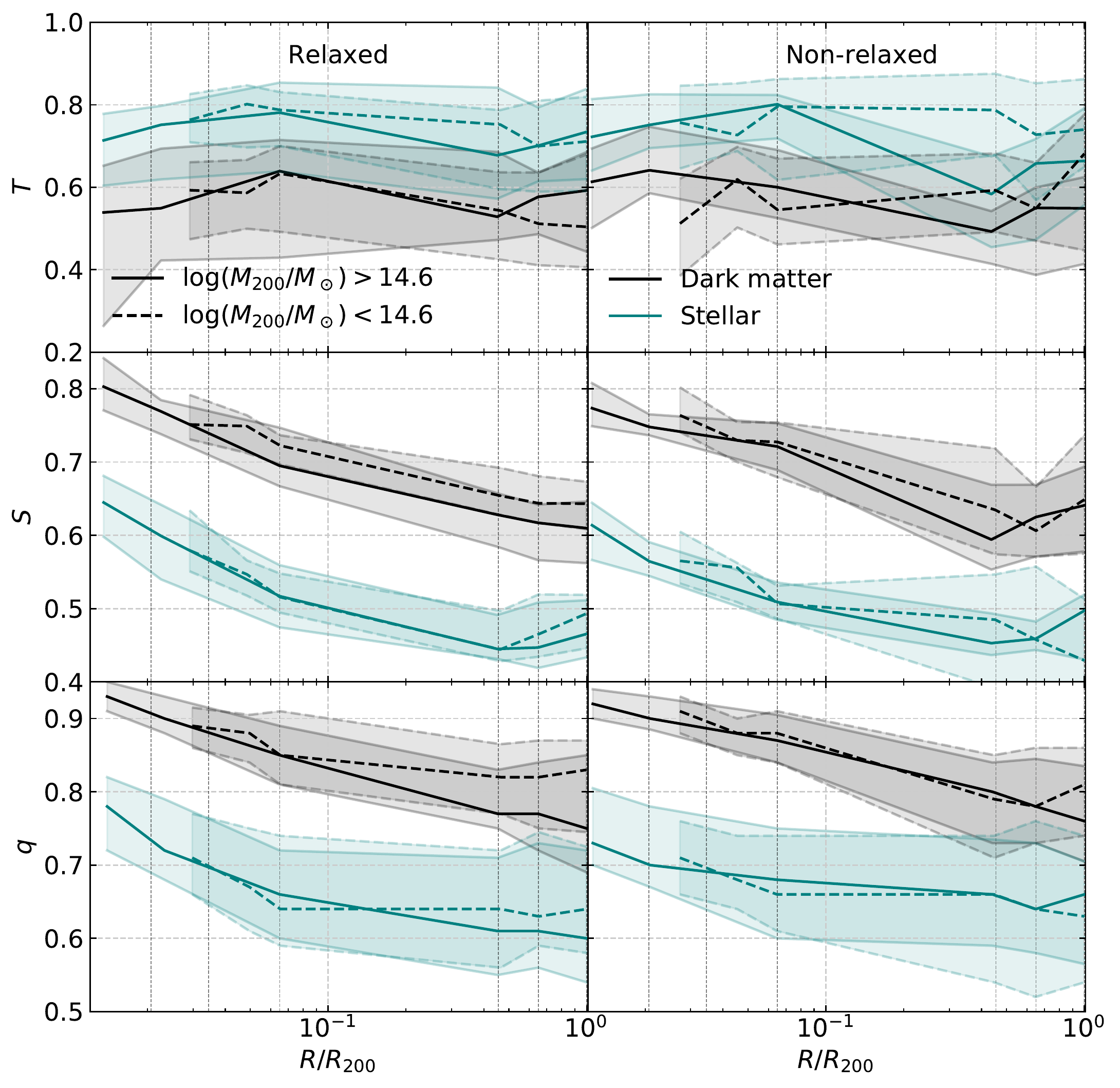}
    \caption{Median values of the shape parameters derived for the total dark matter particle  (black lines) and the stellar (green lines) distributions within the considered radius scaled with $R_{200}$. Shaded regions enclose from 25$^{\rm th}$ to 75$^{\rm th}$ percentiles. The total cluster sample is split according to the cluster mass, $M_{200}$, and relaxation state. Higher and lower mass cluster relations are plotted in solid and dashed lines, respectively. Relaxed (left panel) and non-relaxed (right panel) clusters are classified according to $M_\text{sat}/M_\text{BCG}$. Vertical lines correspond to the median scaled radii for the total sample of clusters listed in Table \ref{tab:radius}.}
    \label{fig:DMstars}
\end{figure*}

\section{Relation between stellar and dark matter particle distribution}
\label{A:stars}

Here we study how the shape parameters determined according to the stellar distribution are related to the derived from the total dark matter particle distribution. We show in Fig. \ref{fig:DMstars} the median stellar and dark matter shape parameters at different radii for relaxed and non-relaxed clusters and split according the median $\log(M_{200})$ mass. As we can see, the stellar distribution is on average
more prolate and less spherical than the dark matter distribution. This result is in agreement with previous studies \citep{Tenneti2014,Velliscig2015}.

 For high-mass clusters, within the BCG region, median stellar and dark matter distribution triaxiality tend to grow with radius, i.e. BCGs are more prolate at the outskirts. At this region both distributions are more prolate than within $R_{1000}$, where the median triaxility values peak to a minimum and starts to grow again through the outskirts. On the other hand, low-mass clusters show in general a mostly constant median triaxiality with radius. 

As mentioned, for the dark matter distribution, high-mass clusters tend to be less spherical, specially relaxed clusters. This trend is not observed in the stellar distribution, for which low and high-mass clusters show similar median sphericity values for all radius. Stellar and dark matter distributions within the innermost physical radius, 30kpc, are more spherical for high mass clusters, suggesting that BCGs located in these clusters up to 30kpc, are rounder. On the other hand the stellar distribution at the scaled radius $0.1R_{500}$ are in agreement for low- and high-mass clusters. 

Finally we study the alignment between the stellar and the dark matter SMA at each radius. We show the results for the different samples in Fig. \ref{fig:tStarsDM}. In general, the DM SMA orientation is well constrained by the stellar distribution for all the samples with a median misalignment values of $\lesssim 10^\circ$. The orientation is better constrained for high-mass relaxed clusters, with a median angle of $\lesssim 4^\circ$ within 30kpc $< R < R_{200}$. The highest misalignment is observed at $R_{200}$ for all the cluster samples.

\section{3D Relation between galaxy members and DM particle distributions}
\label{A:gxs}

\begin{figure*}
    \includegraphics[scale=0.6]{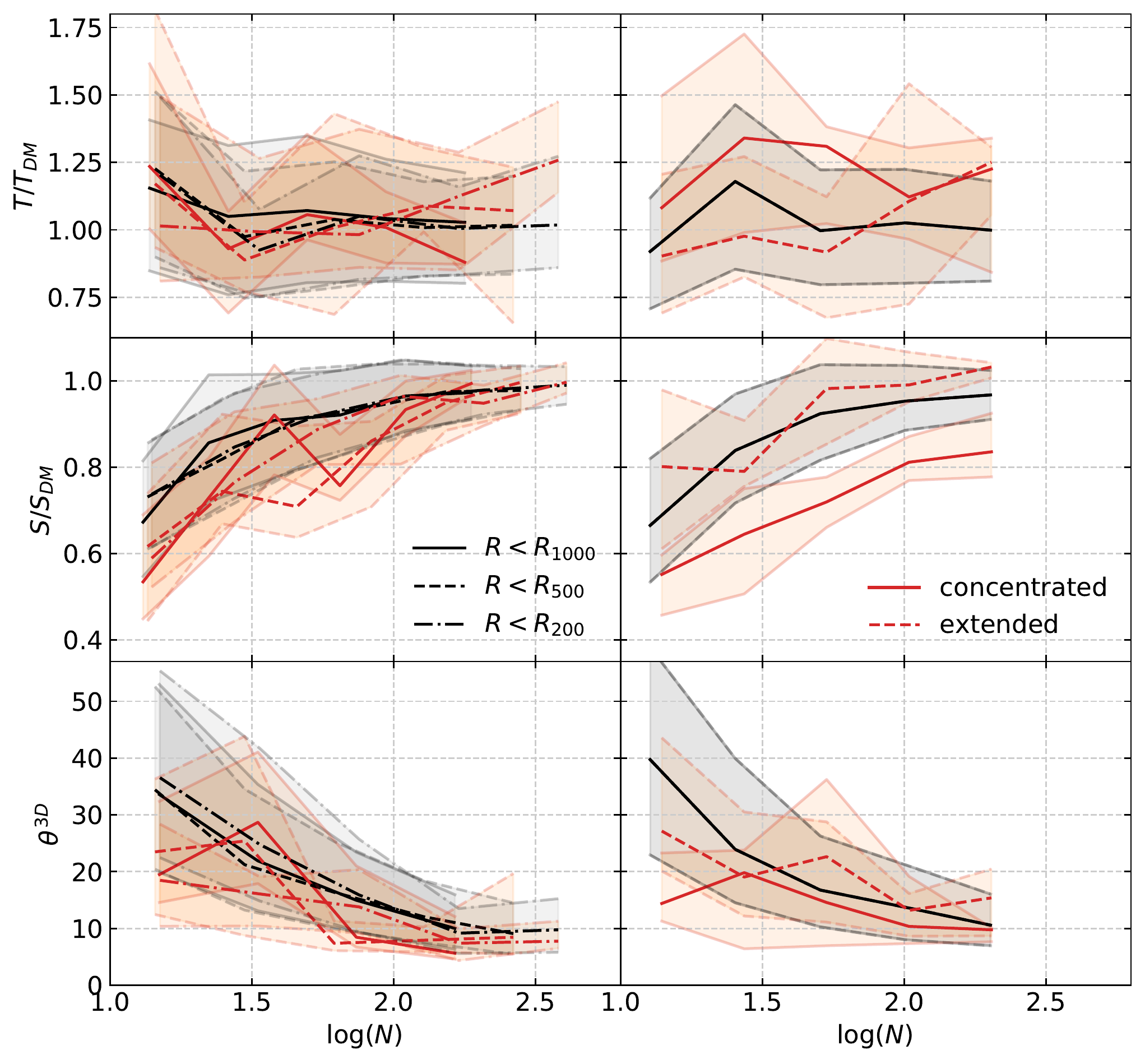}
    \caption{Left panels: Red (black) lines correspond to median 3D shape parameters and misalignment angle between the galaxy member distribution (randomly selected DM particles) and the total DM particle distribution in bins of $\log(N)$, where $N$ is the number of subhalos (particles) within $R_{1000}$, $R_{500}$  and $R_{200}$.  Shaded regions enclose from 25th to 75th percentiles. Right panels: Same as in the left panels but taking the subhalos and particles within $R_{200}$ and considering the subhalos classified as \textit{concentrated} and \textit{extended}. }
    \label{fig:shape_gx_dm}
\end{figure*}

Here we show the results of the comparison between the total DM and the galaxy member distributions. Galaxy members are identified and classified as \textit{concentrated} and \textit{extended} as stated in \ref{subsec:members}. For the comparison we use the same approach as addressed in the subsection \ref{subsec:gx}, by computing the shape parameters selecting a random number of DM particles, according to the number of galaxies identified in each cluster.

Results for the 3D shape parameters are shown in Fig. \ref{fig:shape_gx_dm}. The sphericity ratio is on average
lower for the galaxy distribution than for the randomly DM selected realisations, regardless of the number of subhalos considered.
Triaxiality, on the other hand, is on average 
in agreement with the total dark matter distribution, except for higher mass clusters (higher $N$) where the galaxy distribution up to $R_{200}$ tends to be more prolate. There is also a better alignment between the subhalos and the total DM distribution, than the one obtained according to the randomly selected sample, specially when considering the tracers up to $R_{200}$. More \textit{concentrated} galaxies follow a less spherical and more prolate distribution, and are better aligned with the total DM SMA than \textit{extended} galaxies. This result is in agreement with the projected shape distributions discussed in \ref{subsec:gx}.


\bsp	
\label{lastpage}
\end{document}